\documentclass[11pt]{article}
\usepackage{fullpage}
\usepackage{wustyle}
\usepackage{thmtools}
\usepackage{thm-restate}
\usepackage{verbatim}

\usepackage[top=2.54cm,left=2.5cm,right=2.5cm,bottom=2.54cm]{geometry}

\usepackage[square]{natbib}
\usepackage[textsize=tiny]{todonotes}

\newcommand{\abs}[1]{\left\vert#1\right\vert}
\newcommand{\norm}[2]{\left\Vert#1\right\Vert_{#2}}
\newcommand{\inn}[1]{\left\langle#1\right\rangle}
\newcommand{\ep}{{\epsilon}}
\newcommand{\spg}[1]{\text{span}\{#1\}}
\newcommand{\rloc}{\mathcal{R}_{\text{loc}}}
\let\leq=\leqslant 
\let\geq=\geqslant %

\begin{document}

\title{The Local Landscape of Phase Retrieval Under Limited Samples}

\author{
  Kaizhao Liu$^*$\\
  School of Mathematical Sciences\\
  Peking University\\
  \texttt{mrzt@stu.pku.edu.cn}
  \and
  Zihao Wang\thanks{Equal contribution}\\
  Department of Mathematics\\
  Stanford University\\
  \texttt{zihaow@stanford.edu}
  \and
  Lei Wu\\
  School of Mathematical Sciences\\
  Center for Machine Learning Research\\
  Peking University\\
  \texttt{leiwu@math.pku.edu.cn}
}
\maketitle

\begin{abstract}
In this paper, we present a fine-grained analysis of the local landscape of phase retrieval  under the regime of limited samples. Specifically, we aim  to ascertain the minimal sample size required to guarantee a benign local landscape  surrounding global minima in high dimensions.
Let $n$ and $d$ denote the sample size and input dimension, respectively. 
We first explore the  local convexity and establish that when $n=o(d\log d)$, for almost every fixed point in the local ball, the Hessian matrix has negative eigenvalues, provided $d$ is sufficiently large. 
We next consider the one-point convexity and show that, as long as $n=\omega(d)$, with high probability, the landscape is one-point strongly convex in the local  annulus: $\{w\in\mathbb{R}^d: o_d(1)\leqslant \|w-w^*\|\leqslant c\}$, where $w^*$ is the ground truth and $c$ is an absolute constant. This implies that gradient descent, initialized from any point in this domain, can converge to an $o_d(1)$-loss solution exponentially fast. Furthermore, we show that when $n=o(d\log d)$, there is a radius of $\widetilde\Theta\left(\sqrt{1/d}\right)$ such that  one-point convexity breaks down in the corresponding smaller local ball. This  indicates an impossibility to establish a convergence to the exact $w^*$ for gradient descent under limited samples by relying solely on  one-point convexity.



\end{abstract}

\section{Introduction}

Non-convex optimization arises in many  applications, including matrix decomposition~\citep{bhojanapalli2016global,zhao2015a,chi2019nonconvex,ge2018matrix,chen2020noisy,chi2019nonconvex}, tensor decomposition~\citep{ge2015escaping,fu2020computing}, linear integer programming~\citep{Genova2011LINEARIP}, and phase retrieval ~\citep{waldspurger2013phase,candes2015matrix,candes2015,Netrapalli2015phase,Sun_2017}. The success of deep learning~\citep{alex2012,goodfellow2016deep} has particularly underscored the importance of non-convex optimization.  Among all approaches for tackling these problems, gradient-based methods are especially favored in practice due to their straightforward implementation and versatility across a broad spectrum of problems. Therefore, a thorough understanding of how gradient-based algorithms, including gradient descent and its variations, perform in the realm of non-convex optimization is crucial.


In this paper, we focus on   the following non-convex  problem
\begin{equation}\label{eq:loss}
    \min_{w} L(w):=\frac{1}{4n}\sum_{i=1}^n \left((w^{\top}x_i)^2-y_i^2\right)^2,
\end{equation}
where $x_i\in \RR^d$, $i=1,\dots,n$ are the input samples, $y_i={w^*}^{\top}x_i$  are the corresponding labels, and  $w^*\in\RR^d$  denotes the ground truth.  
Moreover, throughout this paper, we make the following assumption: 
\begin{assumption}\label{assumption: 1}
    $x_i\stackrel{iid}{\sim}\cN(0,I_d)$ for $i=1,2,\dots,n$ and $\|w^*\|=1$.
\end{assumption}

 On the one hand,  Problem \eqref{eq:loss}  is exactly the real phase retrieval problem with the least square formulation~\citep{dong2023phase}: one concerns how to recover an unknown signal $w^*\in\RR^d$ from a series of magnitude-only measurements
\begin{equation}\label{eqn: x3}
y_i=\abs{w^{*\top}x_i}, \quad i=1,\dots,n.
\end{equation}
Solving the problem \eqref{eqn: x3} through minimizing the least-square loss gives Problem \eqref{eq:loss}. Phase retrieval is important for many applications in physics and engineering~\citep{shechtman2014phase,elser2018benchmark,Hohage_2019,dong2023phase}. 

On the other hand, Problem \eqref{eq:loss} can also be viewed as learning a single neuron with the quadratic activation function $\sigma(z)=z^2$. This  is often adopted as a pedagogical example to understand the non-convex optimization involved in training neural networks~\citep{pmlr-v80-du18a,mannelli2020optimization,yehudai2022learning,frei2020agnostic,wu2022learning,Mignacco_2021}.

\paragraph*{Gradient Descent.} Numerous methods have been proposed to solve Problem~\eqref{eq:loss} by leveraging its particular structure, such as spectral initialization~\citep{Netrapalli2015phase} and approximate message passing \citep{schniter2014compressive}. However, in practice,  plain gradient descent $w_{t+1} = w_t - \eta \nabla L(w_t)$ with random initialization also performs surprisingly well~\citep{Chen_2019}. This naturally raises the question of why gradient descent  works so well despite the non-convexity. 

On the one hand, one can directly analyze the trajectory of gradient descent. Specifically,~\citet{Chen_2019} adopted this approach and proved that when $n=\Omega(d\log^{13} d)$, gradient descent can converge to a solution with $\varepsilon$ error  in $\cO(\log d + \log(1/\varepsilon))$ iterations. However, the requirement on sample size $n$ is far from being optimal since numerical experiments suggest that $n=\Theta(d)$ might be sufficient for the success of gradient descent \citep{lenka2020}.

On the other hand, one can  characterize the landscape of $L(\cdot)$, aiming  to show that the loss landscape has certain benign properties, such as strict saddle property and local strong convexity~\citep{Sun_2017,cai2021perturbed,cai2022quotient,cai2022solving,Cai_2023}. These benign landscape properties can imply a global convergence of (perturbed) gradient descent in polynomial time~\citep{jin2017escape}. 


\paragraph*{Landscape.} In this paper, we take the landscape approach. 
Before analyzing the empirical landscape $L(\cdot)$, it is helpful to first take a glimpse of the population landscape 
\begin{equation}
\bar{L}(w):=\frac 1 4\EE_{x}\left[(w^\top x)^2-({w^*}^\top x)^2\right]=\frac{1}{4}\left(3\|w\|^4+3\|w^*\|^4-2\|w\|^2-4(w^{\top}w^*)^2\right).
\end{equation}
A simple calculation gives 
\begin{equation*}
\begin{aligned}
\nabla \bar{L}(w) &= (3\|w\|^2 -1)w - 2 (w^{\top}w^*) w^*\\ 
\nabla^2 \bar{L}(w) &= 6ww^{\top}-2 w^*w^{*\top} + (3\|w\|^2-1) I.
\end{aligned}
\end{equation*}
It is easy to verify that  the critical points of population landscape $\bar{L}(\cdot)$ are given by 
\begin{itemize}
\item global maxima: $w=0$;
\item saddle points: $\|w\|^2=1/3$ and $w\perp w^*$; 
\item global minima: $w=\pm w^*$.
\end{itemize}
Moreover, it is easy to verify  that the population landscape is benign in the following sense.
\begin{property}\label{property: landscape}
The landscape has no spurious local minima, all saddle points are strict (i.e., the Hessian matrix has negative eigenvalues), and the local landscape around global minima is strongly convex, which, consequently, implies one-point strong convexity (see Definition \ref{def: one-point}). 
\end{property}
The above  property of landscape implies that (perturbed) gradient descent with random initialization can find a global minimum in polynomial time. In addition,  if the sample size $n$ is large enough, 
it is not surprising that Property \ref{property: landscape} also holds for the empirical landscape $L(\cdot)$. However, the challenging question is determining the smallest $n$ required to enable these benign properties.


\citet{Sun_2017} showed $n= \Omega(d\log^3 d)$ suffices for establishing Property \ref{property: landscape}. Further refinements by  \citet{Cai_2023}  show that : 1) all saddle points distant from global minima are strict when $n=\Omega(d)$ and 2) the local landscape in the vicinity of global minima is strongly convex if $n=\Omega(d\log d)$. Nonetheless,  the preceding  sample complexities may not be optimal for several reasons.
 First, from an information-theoretical perspective, it is possible to recover the ground truth  using only $n=\Omega(d)$ samples and indeed, prior works have designed other algorithms to achieve this feat~\citep{chen2016,cai2021perturbed,cai2022quotient,cai2022solving}. Second,
empirical works have suggested that plain gradient descent with random initialization is effective in solving Problem \eqref{eq:loss} even when $n=\Theta(d)$. For instance, \citet{lenka2020} employed experiments and non-rigorous replica methods to hypothesize that $n=13.8d$ may be adequate. Consequently, it raises an intriguing question:
\begin{center}
    \textit{Can we establish the benign property of landscape for the non-convex optimization problem \eqref{eq:loss} when  $n=o(d\log d)$ or even $\cO(d)$?}
\end{center}


\paragraph*{Our Contributions.}
As mentioned above, \citet{Cai_2023} already demonstrated  if $n=\Omega(d)$, outside a local region, there are no spurious local minima and all saddle points are strict. 
 Therefore,  in the current work, we narrow our focus  to the  landscape in the vicinity of global minima. 

\begin{figure}[!h]
    \centering
    \includegraphics[width=0.5\textwidth]{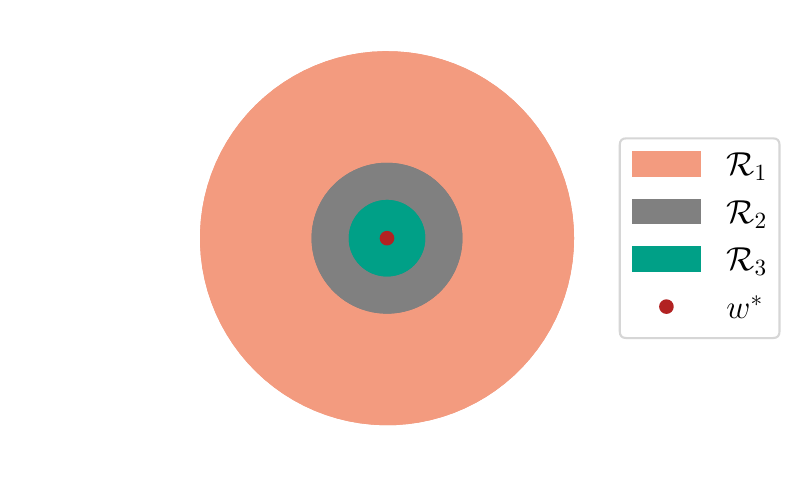}
    \vspace*{-1em}
    \caption{A schematic illustration of our characterizations of the local landscape. We prove that when $n\in [\omega(d),o(d\log d)]$,  1) the landscape is non-convex but one-point strongly convex in $\cR_1$; 2) the landscape is neither convex nor one-point convex in $\cR_3$.
    }
    \label{fig:result}
\end{figure}

 To clearly state our contribution, we decompose the local region into three subdomains:
\begin{align*}
\cR_1&=\{w\in\RR^d: r_{1,d}\leq \|w-w^*\|\leq c\}\\ 
\cR_2&=\{w\in\RR^d: r_{2,d}\leq \|w-w^*\|\leq r_{1,d}\}\\
\cR_3&=\{w\in\RR^d: \|w-w^*\|\leq r_{2,d}\},
\end{align*}
where $c$ is a small absolute constant, $r_{1,d}$ is an $o_d(1)$ quantity and $r_{2,d}=\Theta\left(\sqrt{\frac{\log n}{d}}\right)$. See Figure \ref{fig:result} for a schematic illustration. Let $\overline{\cR}=\cR_1\cup \cR_2\cup \cR_3$ denote the entire local domain.  

\begin{itemize}
\item In Section \ref{sec:negative}, we examine the convexity of local landscape. 
We prove that when $n=o(d\log d)$, for almost every fixed point in $\overline{\cR}$, the Hessian matrix at that point, with high probability, has negative eigenvalues when $d$ is sufficiently large. Consequently, local landscape must be highly non-convex when $n=o(d\log d)$.

\item In Section \ref{sec: positive}, we investigate the one-point convexity of local landscape. On the positive side, we establish that $L(\cdot)$ is one-point strongly convex in $\cR_1$ if $n=\omega(d)$, which implies a local convergence to $o_d(1)$-loss solutions. On the negative side, we establish that when $n=o(d\log d)$, the one-point convexity breaks down in $\cR_3$. This indicates that with limited samples, it is impossible to guarantee convergence to the exact global minima by merely utilizing local one-point convexity.
\end{itemize} 
For a better understanding, we provide a summary of our results in Figure \ref{fig:result}.

Furthermore, to establish the aforementioned negative results, we introduce an ``add-one trick'' to disentangle the dependence when estimating the summation of  dependent random variables. For more details, we refer to Section \ref{sec: add-one}. This technique might be of independent interest.








\subsection{Notations}

Let  $[k]=\{1,2,\dots,k\}$ for any $k\in \NN$.
For a vector $v$, denote by $\|v\|:=(\sum_i |v_i|^2)^{1/2}$ the $\ell^2$ norm.
For a matrix $A$, denote by $\|A\|$ and $\|A\|_F$ the spectral norm and Frobenius norm, respectively. Let $\SS^{d-1}=\{x\in\RR^{d-1}:\|x\|=1\}$ denote the unit sphere. Given  $S=\{v_1,\dots,v_k\}$, denote by $\spg{S}$  the linear span of $S$ and $\spg{S}^{\perp}$ the orthogonal complement. 

hroughout this paper, we use $C$ and $c$ to denote sufficiently large and sufficiently small absolute positive constants, respectively. Their values may vary from line to line. We also use the standard big-O notations:  $\cO(\cdot)$, $\Theta(\cdot)$,$\Omega(\cdot)$ to only hide absolute constants. In addition, we use $\widetilde{\cO}$, $\widetilde{\Theta}$ and $\widetilde{\Omega}$ to hide logarithmic terms, e.g., $\cO(d\log d) = \widetilde{\cO}(d)$.
We also use $\omega_d(\cdot)$ and $o_d(\cdot)$ notations. Here,
$f(d) = \omega_d(g(d))$ means  $f(d)/g(d)\to\infty$ as $d\to\infty$ and $f(d) = o_d(g(d))$ means  $f(d)/g(d)\to 0$ as $d\to\infty$. We sometimes omit the subscript for simplicity when it is clear from the context.

\section{The Break of Local Convexity}\label{sec:negative}

For any function $f\in C^2(\cD)$ defined in a convex domain $\cD$, the followings are equivalent:
\begin{itemize}
\item $f$ is convex in $\cD$.
\item $\nabla^2 f(x)\succeq 0$, $\forall x\in \cD$.
\end{itemize}
Therefore, to examine the local convexity of $L(\cdot)$, we can check the eigenvalues of its Hessian matrix. 
The Hessian of $L(\cdot)$ is given by 
\begin{equation}\label{eq:hessian-matrix}
    \nabla^2 L(w) = \frac{1}{n}\sum_{i=1}^{n}(3(w^{\top}x_i)^2-(w^{*\top}x_i)^2)x_ix_i^\top.
\end{equation}
We aim to estimate the smallest eigenvalue of $\nabla^2 L(w)$ in the local region around $w^*$. To facilitate our statement, we introduce several additional notations. 
Let $\cP^{\perp} := (I-w^*w^{*\top})$ denote the projection operator onto the orthogonal complement.
Moreover, for any $w\neq w^*$, the direction of $\cP^{\perp}w$ is denoted by $w^{\perp}:=\frac{\cP^{\perp}w}{\norm{\cP^{\perp}w}{}}$. In addition, our local region is defined as
\begin{equation}\label{eqn: local}
\rloc:=\left\{w=\alpha w^*+\beta w^{\perp}: |\alpha-1|\leq\frac{1}{3},\beta\in (0,1] \right\},
\end{equation}
where $\alpha=\langle w,w^*\rangle$ and $\beta=\norm{\cP^{\perp}w}{}$ denote the magnitude of the parallel and orthogonal components, respectively. Note that there are two equivalent global minima, $+w^*$ and $-w^*$. Without loss of generality, we only consider the global minimum $+w^*$ in \eqref{eqn: local} for simplicity.







The following theorem demonstrates that the local landscape is non-convex when $n=o(d\log d)$.

\begin{restatable}{theorem}{mainthmone}\label{thm:local_non_convex}
For any fixed $w\in \rloc$ with $\beta=\|\cP^{\perp}w\|>0$, if  $d\geq C$ and $n\geq Ce^{C/\beta^2}$, then w.p.~at least $1-Ce^{-cd}-C/n-e^{-ce^{-C/\beta^2}\sqrt{n}}$, it holds that
    \begin{equation}\label{eqn: x15}
    \min_{u\in\SS^{d-1}} u^{\top}\nabla^2 L(w) u \leq -c\beta^2\frac{d\log n}{n}+C .
    \end{equation}
\end{restatable}

\begin{corollary}\label{corollary: 1}
When $n\geq d$, $n=o(d\log d)$, and $d\to \infty$,  it holds with probability approaching $1$ that
\begin{equation}
\min_{u\in\SS^{d-1}} u^{\top}\nabla^2 L(w) u \rightarrow-\infty.
\end{equation}
\end{corollary}
\begin{proof}
Let $\gamma_d=n/(d\log d)$. Then, $ \gamma_d \rightarrow 0$ as $d\rightarrow +\infty$ and consequently,
$
\frac{d\log n}{n} \geq  \frac{1}{\gamma_d}\rightarrow+\infty .
$
Plugging it into \eqref{eqn: x15}  completes the proof.
\end{proof}

Thus we establish that the local landscape is non-convex if the sample size $n$ is only $o(d\log d)$. Moreover, Theorem \ref{thm:local_non_convex}  implies that the non-convexity becomes stronger as $d$ grows under the proportional scaling $n/d=\gamma$ where $\gamma$ is a constant.

It is worth noting that the requirement $\beta>0$ and the dependence of $\beta$ in Theorem \ref{thm:local_non_convex} are unavoidable. Consider the case of $\beta=0$.  
Plugging $w=\alpha w^*$ into \eqref{eq:hessian-matrix} gives 
\[
    \nabla^2 L(\alpha w^*) = (3\alpha^2-1)\frac{1}{n}\sum_{i=1}^n ({w^*}^\top x_i)^2x_ix_i^\top.
\]
This implies that the Hessian matrix is always semi-positive definite whenever $\beta=0$ and $|\alpha|\geq 1/\sqrt{3}$. 
To obtain a better understanding of the influence of $\beta$, let us consider the proportional scaling $\gamma=n/d$. If we want the upper bound \eqref{eqn: x15} to be negative, we need
\begin{equation}\label{eqn: x16}
  d\geq d_{\beta,\gamma}:=\frac{1}{\gamma}\exp(C \frac{\gamma}{\beta^2}),
\end{equation}
where $C$ is an absolute constant. We can see  that $d_{\beta,\gamma}$ depends on $\frac{1}{\beta}$ in a super exponential manner. Consequently,  to observe negative curvatures at an arbitrary fixed point, the required dimensionality is astronomical. For example, when $\gamma=2,\beta=0.1$, we have $d_{\beta, \gamma}=0.5\exp(200C)$. This indicates that Theorem \ref{thm:local_non_convex} is effectively asymptotic, which is expected since it guarantees the existence of negative curvatures at an arbitrarily fixed point.

In the following, we further consider  the worst-case situation: whether there exists a point in  $\rloc$ where the Hessian matrix has negative eigenvalues.



\begin{restatable}{theorem}{mainthmfour}\label{thm:improved_local_non_convex}
When $n,d\geq C$ and $\gamma_{n,d}:=C\sqrt{\frac{\log n}{d}}$,
w.p.~at least $1-Ce^{-c\sqrt{d}}-Ce^{-c\log n}$, we have
\begin{equation}\label{eqn: x25}
    \min_{u\in\SS^{d-1},\|w-w^*\|\leq \gamma_{n,d}} u^{\top}\nabla^2 L(w) u \leq -c\frac{d\log n}{n}+C .
    \end{equation}
\end{restatable}
 
Analogous to Corollary \ref{corollary: 1}, 
the above theorem indicates that the local landscape becomes non-convex when $d$ is sufficiently large under the conditions $n\geq d$, $n=o(d\log d)$. Moreover, following the derivation of \eqref{eqn: x16}, 
the condition on $d$ here is independent of $\beta$, which contrasts with the requirement in Theorem \ref{thm:local_non_convex}.

\paragraph{Numerical Experiments.} 
 To validate the findings in Theorem \ref{thm:improved_local_non_convex},  we quantify the non-convexity of local landscape using the following quantity
\begin{equation}\label{eq:minhessian}
    q_r(d):=\min_{u\in \SS^{d-1},\|w-w^*\|\leq r}u^{\top}\nabla^2 L(w)u.
\end{equation}
We will examine how $q_r(d)$ changes with increasing $d$ under the proportional scaling $n/d=\gamma$.  Note that theorem \ref{thm:improved_local_non_convex} shows that $q_r(d)\leq -\frac{c\log(\gamma d)}{\gamma}+C$, suggesting that $q_r(d)$ decreases as $d$ grows.

\begin{figure}[!h]
    \centering
    \includegraphics[width=0.4\textwidth]{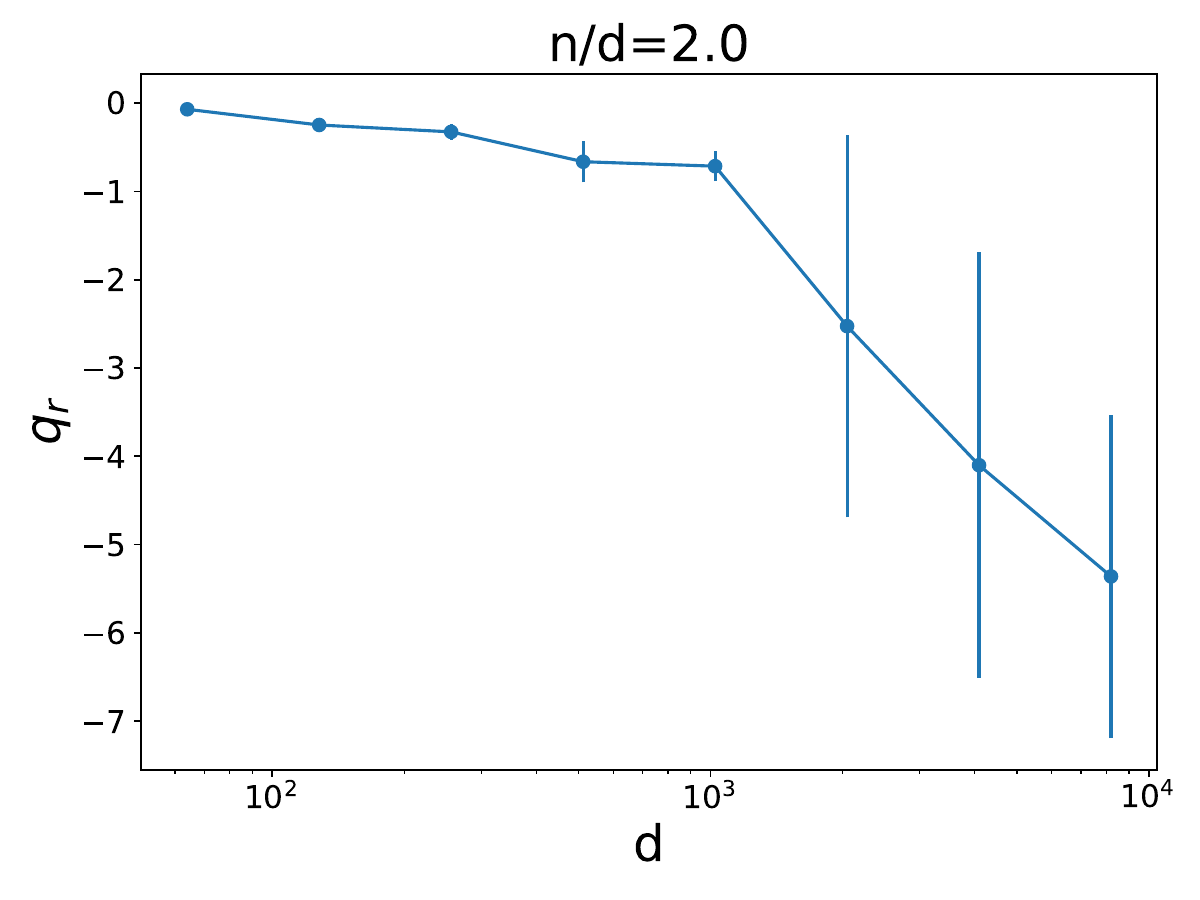}
    \includegraphics[width=0.4\textwidth]{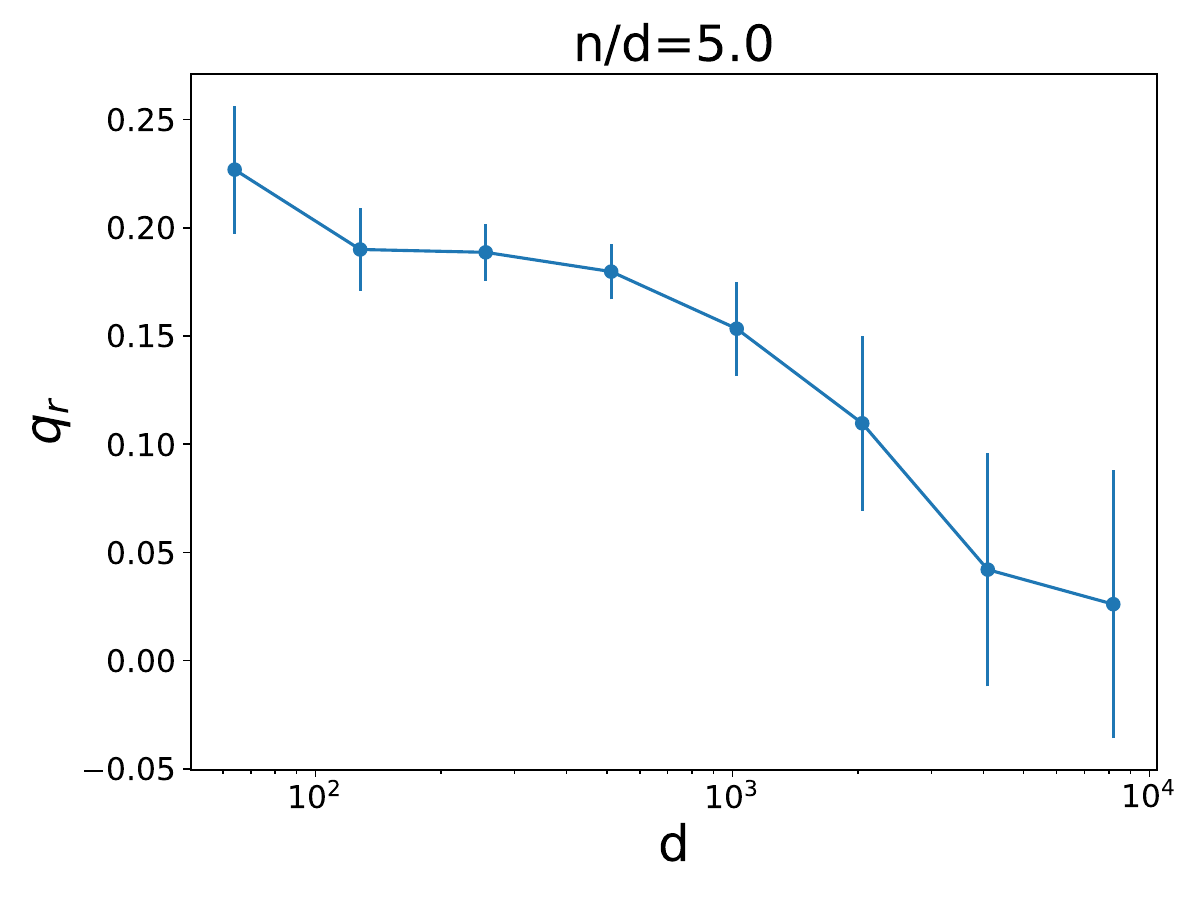}
    \vspace*{-.5em}
    \caption{$q_r(d)$ with $r=0.1$ for $n/d=2$ and $n/d=5$. The results are averaged over 10 random seeds. The optimization is executed with the Adam optimizer with hyperparameters $(\beta_1,\beta_2)=(0.9,0.999)$. After each optimization step, a projection onto the constrained domain is performed. The learning rate schedule involves using $0.001$ for the first 200 steps, followed by $0.0005$ for the next 200 steps, and concluding with $0.0003$ for the last 600 steps. At initialization, $u$ and $w$ are uniformly sampled from $\SS^{d-1}$ and $\{w\in\RR^d:\|w-w^*\|= r\}$, respectively. Note that the population landscape is strongly convex within $\{w\in\RR^d: \norm{w-w^*}{} \leq 0.1\}$.}
    \label{fig:opt}
\end{figure}

In experiments, we solve the optimization problem \eqref{eq:minhessian} by using Adam optimizer~\citep{kingma2014adam} with projection to the constraint domain in each step. The results are shown in Figure \ref{fig:opt}. We can see very clearly that $q_r(d)$ decreases as $d$ grows. This is consistent with our theoretical findings in Theorem \ref{thm:improved_local_non_convex}, i.e., the ``non-convexity'' of local landscape becomes stronger for larger $d$. Specifically, we can see when $n/d=5$, the local landscape becomes non-convex as long as $d$ is larger than $10,000$.

\section{Local One-Point Convexity}\label{sec: positive}
Beyond the classical convexity condition, another popular sufficient condition for establishing local convergence is the one-point strong convexity. 
\begin{definition}[One-Point Strong Convexity]\label{def: one-point}
$L(\cdot)$ is said to be local one-point strongly convex with respect to $w^*$ if  for any $w$ satisfying $\|w-w^*\|\leq c$ it holds that 
\begin{equation}\label{eqn: one-point-convex}
\langle \nabla L(w),w-w^*\rangle \geq c \|w-w^*\|^2.
\end{equation}
\end{definition}
One-point strong convexity is weaker than strong convexity as the latter implies the former but the reverse is not true; see Figure 4 in~\citet{li2017convergence} for a concrete example. One-point strong convexity ensures that negative  gradient points toward a good direction. Specifically,  for gradient flow
$
\dot{w}_t = -\nabla L(w_t),
$
under the one-point strong convexity, we have
\begin{equation}\label{eqn: one-point-gf}
\frac{\dd}{\dd t}\norm{w_t-w^*}{}^2 = 2 \langle w_t-w^*, \dot{w}_t\rangle
=-2\langle w_t-w^*,\nabla L(w_t) \rangle \leq -2c\norm{w_t-w^*}{}^2,
\end{equation}
thereby $\|w_t-w^*\|=\cO(e^{-ct})$. 
This implies that one-point strong convexity can guarantee exponential fast convergence of gradient flow.

One-point strong convexity has been widely utilized in non-convex optimization to establish convergence, including  learning a single neuron~\citep{yehudai2022learning,wu2022learning}, training neural networks~\citep{li2017convergence,kleinberg2018alternative}, and phase retrieval~\citep{candes2015,chen2016}.  It is also worth noting that one-point strong convexity  is equivalent to the quasi-strong monotonicity for the gradient operator: $w \mapsto \nabla L(w)$, which has been  widely used in studying variational inequality problems~\citep{harker1990finite,sadiev2023high}.


Then, it is natural to ask what is the smallest sample size to ensure local one-point convexity for phase retrieval. The following theorem provides a positive answer to this question. 
\begin{restatable}[Positive Result]{theorem}{mainthmthree}\label{thm: one point convex positive}
   For any $d,t\geq C$,  if $n\geq Ct^2d$, then  w.p.~at least $p_{d}:=1-C\frac{e^{\frac{t^2}{2}}}{td}-Ce^{-cd}$, we have
    \[
    \inf_{w\in \cD_{t,d}}\frac{\langle \nabla L(w),w-w^*\rangle}{\norm{w-w^*}{}^2} \geq c, 
    \]
    where $\cD_{t,d}$ is a local annulus given by $\cD_{t,d}:=\{w\in\RR^d:Ct^3e^{-\frac{t^2}{2}}\leq \norm{w-w^*}{}^2\leq c \}$.
\end{restatable}

When $t^2\in [\omega(1),o(\log d)]$, we have  $t^3e^{-\frac{t^2}{2}}=o_d(1)$ and the probability $p_d \to 1$ as $d\to \infty$. Therefore, when $n\in [\omega(d),o(d\log d)]$, the landscape is one-point strongly convex in the local annulus $\{w\in\RR^d: o_d(1)\leq \norm{w-w^*}{}^2\leq c\}$.
This implies that the local landscape is somewhat benign  as long as $n=\omega(d)$, despite being non-convex according to Theorem~\ref{thm:local_non_convex}. 
In a stark contrast, the benign property (strong convexity) of local landscape established in~\citet{Cai_2023} requires $n=\Omega(d\log d)$.

It is worth noting that~\citet{Cai_2023} showed when $n=\Omega(d)$, outside a local region, all saddle points are strict and there are no spurious local minima. Applying the results of studying saddle-point escape~\citep{jin2017escape}, the preceding landscape properties suggest that (perturbed) gradient descent with random initialization can enter the local region in polynomial time as long as $n=\cO(d)$. Our result (Theorem \ref{thm: one point convex positive})  further shows that the local convergence to an $o_d(1)$-loss solution only needs $n=\omega(d)$. These results together strongly suggest that  $n=\omega(d)$ should suffice for (perturbed) gradient descent with random initialization to find an $o_d(1)$-loss solution in polynomial time. In contrast, previous works need more samples to guarantee global convergence for gradient descent with random initialization. Specifically, \citet{arous2021online} showed that online  stochastic gradient descent can reach  an $o_d(1)$-loss solution when $n=\omega(d\log^2 d)$. \citet{Chen_2019} proved that when $n=\Omega(d\log^{13} d)$, gradient descent converges to an $\varepsilon$-loss solution in $\cO(\log d+\log(1/\varepsilon))$ iterations. However, to make this  claim fully rigorous needs a careful characterization of how strict  saddle points are and how to lower bound the gradient  norm for non-saddle points. We leave this interesting question to future work.


Then, a natural question is: what about the landscape of the local region within the $o_d(1)$ radius. The following theorem provides a negative result.
\begin{restatable}[Negative Result]{theorem}{mainthmtwo}\label{thm: non one point convex}
When $n,d\geq C$ and $\gamma_{n,d}:=C\sqrt{\frac{\log n}{d}}$,
w.p.~at least $1-Ce^{-c\sqrt{d}}-Ce^{-c\log n}$, we have\[
\min_{\|w-w^*\|\leq \gamma_{n,d}} \frac{\inn{\nabla L(w),w-w^*}}{\norm{w-w^*}{}^2} \leq -c\frac{d\log n}{n} +C .
\]
\end{restatable}

Analogous to Corollary \ref{corollary: 1},  when $n=o(d\log d)$, it holds with probability approaching $1$ as $d\to \infty$ that
\begin{equation}\label{eqn: x11}
\min_{\|w-w^*\|\leq \gamma_{n,d}} \frac{\inn{\nabla L(w),w-w^*}}{\norm{w-w^*}{}^2} \to -\infty.
\end{equation}
This implies that both classical convexity and the local one-point convexity break when $n=o(d\log d)$. The breakdown of the classical convexity is evident because, if the convexity holds, then $\inn{\nabla L(w),w-w^*}\geq 0$ would be true for all $w$ within this region, contradicting Eq.~\eqref{eqn: x11}.

In addition, it is important to note that the locality size $\gamma_{n,d}$ shrinks to zero as $d\to\infty$, provided $n=o(\exp(d))$. This is particularly unexpected given that the Hessian matrix at exactly $w^*$ remains positive definite as long as $n=\Omega(d)$:
\begin{restatable}{lemma}{mainlem}\label{thm: strong-convex}
If $n\geq Cd$, then w.p.~at least $1-\frac{C}{n}-Ce^{-cn}$, we have $\lambda_{\min}(\nabla^2 L(w^*))\geq c>0$.
\end{restatable}
Proof of Lemma \ref{thm: strong-convex} is deferred to  Appendix \ref{sec: proof-hessian}.

\begin{figure}[!h]
    \centering
    \includegraphics[width=0.31\textwidth]{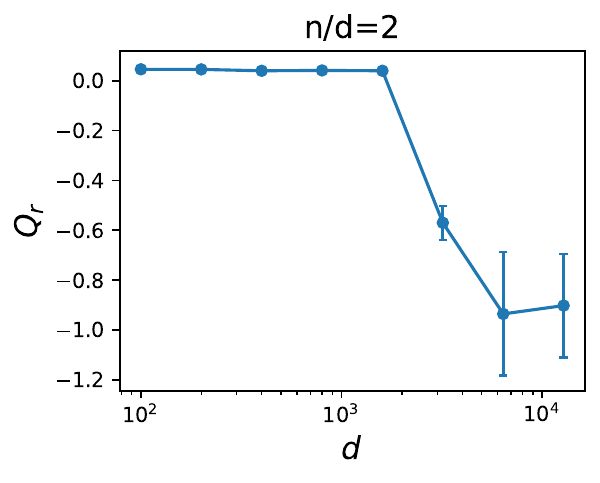}
    \hspace*{-1em}
    \includegraphics[width=0.31\textwidth]{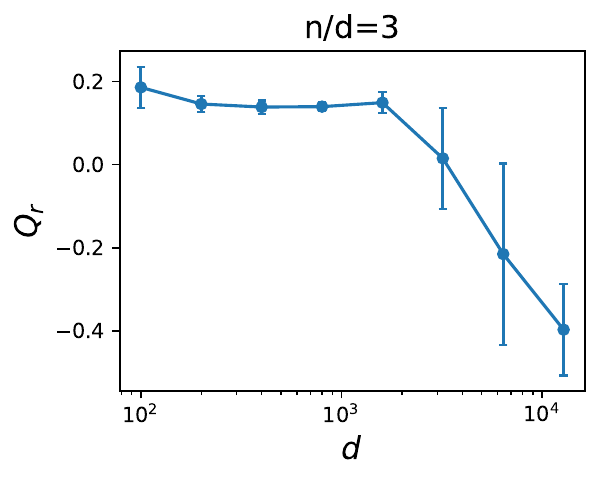}
    \hspace*{-1em}
    \includegraphics[width=0.3\textwidth]{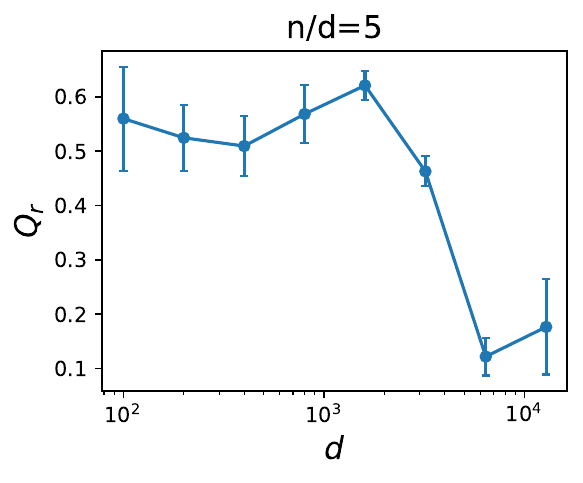}
    \caption{ How $Q_r(d)$ varies with $d$ under the proportional scaling $n/d=2,3$, and $5$. 
    The results are averaged over 10 random seeds. The optimization is executed for $3000$ steps  using the Adam optimizer with hyperparameters $(\beta_1,\beta_2)=(0.9,0.999)$ and learning rate $0.01$. After each optimization step, a projection onto the constrained domain is performed. At initialization,  $w$ is uniformly sampled from $\{w\in\RR^d:\|w-w^*\|= r\}$. In experiments, we set $r=0.1$ and  the population landscape is strongly convex within $\{w\in\RR^d: \norm{w-w^*}{} \leq 0.1\}$.
    }
    \label{fig: one-point}
    \end{figure}

\paragraph*{Numerical Experiments.}
To assess the degree of local one-point convexity, we employed the following metric:
    \begin{align}\label{eqn: 5}
    Q_r(d)  =\min_{\|w-w^*\|\leq r} \frac{\inn{\nabla L(w),w-w^*}}{\norm{w-w^*}{}^2}. 
    \end{align}
    We investigated how $Q_r(d)$ varies with increasing $d$ under the proportional scaling $n/d=\gamma$. The numerical results,  shown in Figure \ref{fig: one-point}, indicate a clear trend: the value of $Q_r(d)$ tends to decrease as $d$ increases. This pattern suggests that under proportional scaling, the local landscape increasingly exhibits one-point non-convexity with higher dimensionality, aligning with  Theorem 3.3. 

    It is worth mentioning that in experiments, we observe that when $\gamma$ is relatively large, such as 5, the optimization problem \eqref{eqn: 5} becomes extremely challenging due to the presence of numerous poor local minima, making it difficult for optimizers to locate  global minima. Different initializations and hyperparameters often lead to different minima. Consequently, in the case of $n/d=5$, the value of $Q_r$ exhibits significant fluctuations as $d$ increases, although the overall trend remains consistent with expectations.

\section{Proofs}
\label{sec: proofs}

\subsection{The Add-One Trick}
\label{sec: add-one}
To prove  the three negative results: Theorem \ref{thm:local_non_convex}, Theorem \ref{thm:improved_local_non_convex}, and \ref{thm: non one point convex}, we need to introduce the following ``add-one trick'' to handle the dependence among summands. 

Let $Z_1,\dots,Z_n$ and $Y_i,\dots, Y_n$ be \iid random variables respectively, and additionally, $(Z_1,\dots, Z_n)$ are independent of $(Y_1,\dots, Y_n)$. Let $J\in \sigma(Y_1,\dots, Y_n)$ be a \textit{random index}, which depends only on $(Y_1,\dots,Y_n)$ and is independent of $(Z_1,\dots,Z_n)$.  For instance, $J=\argmin_{j\in [n]} Y_j$.

Consider the estimation of the following quantity
\begin{align}\label{eqn: o0}
\frac{1}{n}\sum_{i=1}^n f(Z_i, Z_J, Y_i).
\end{align}
We cannot directly apply concentration inequalities  as the index $J$ introduces dependence among summands. To deal with this dependence, we introduce the following \textbf{add-one decomposition}:
\begin{align}\label{eqn: add-one}
\notag \frac{1}{n}\sum_{i=1}^n f(Z_i, Z_J, Y_i) &=\frac{1}{n}\left( f(Z_J, Z_J, Y_J) + \sum_{i\neq J} f(Z_i, Z_J, Y_i)\right)\\ 
&= \underbrace{\frac{1}{n}\left(f(Z_J, Z_J, Y_J) - f(Z_{n+1}, Z_J, Y_J)\right)}_{I_1} + \underbrace{\frac{1}{n}\left(f(Z_{n+1}, Z_J, Y_J) + \sum_{i\neq J} f(Z_i, Z_J, Y_i)\right)}_{I_2},
\end{align}
where $Z_{n+1}$ is an independent copy of $Z_i$. We now proceed to bound $I_1$ and $I_2$ separately.  

For estimating $I_1$, we shall need the following little lemma. 
\begin{lemma}\label{lemma: same-distribution}
Suppose that $J$ is a random variable defined on the index set $[n]$ and $J$ is independent of $Z_1,\dots,Z_n$. Then, we have 
$Z_J\overset{d}{=}Z_i$ for all $i=1,\dots, n$.
\end{lemma}
\begin{proof}
Let $Z$ be an independent copy of $Z_i$ and $\varphi(t):=\EE[e^{it^{\top}Z}]$ be the characteristic function. Then,
$
\EE\left[e^{it^{\top}Z_J}\right]=\EE\left[\EE\left[e^{it^{\top}Z_J}|J\right]\right]=\EE\left[\varphi(t)|J\right] = \varphi(t),
$
where the second step uses the independence between $J$ and $(Z_1,\dots,Z_n)$.  Thus, we complete the proof.
\end{proof}

For bounding $I_2$, we need the following lemma.

\begin{lemma}\label{lemma: add-one}
$f(Z_{n+1}, Z_J, Y_J) + \sum_{i\neq J} f(Z_i, Z_J, Y_i) \overset{d}{=} \sum_{i} f(Z_i, Z_{n+1}, Y_i)$
\end{lemma}
\begin{proof}
Conditioned on $Y_1=y_1,\cdots, Y_n=y_n$ and $J=j$, we have by symmetry that

    \begin{align}\label{eqn: O1}
    \notag f(Z_{n+1}, Z_j, y_j) + \sum_{i\neq j} f(Z_i, Z_j, y_i) &\overset{d}{=} f(Z_j, Z_{n+1}, y_{j}) + \sum_{i\neq j} f(Z_i, Z_{n+1}, y_i)\\ 
    &= \sum_{i=1}^n f(Z_i, Z_{n+1}, y_i),
    \end{align}

where the first step swaps $Z_j$ and $Z_{n+1}$ and this swap does not alter the distribution as $Z_1,\dots,Z_{n+1}$  are \iid random variables. 
Now for any $t\in\RR$, we have the characteristic functions satisfying
\begin{align*}
\EE[e^{it \left(f(Z_{n+1}, Z_J, Y_J) + \sum_{i\neq J} f(Z_i, Z_J, Y_i)\right)}] &= \EE\left[\EE\left[e^{it \left(f(Z_{n+1}, Z_J, Y_J) + \sum_{i\neq J} f(Z_i, Z_J, Y_i)\right)}| J, Y_1,\dots, Y_n\right]\right]\\ 
&=\EE\left[\EE\left[e^{it \left(\sum_{i} f(Z_i, Z_{n+1}, Y_i)\right)}| J, Y_1,\dots, Y_n\right]\right]\\ 
&=\EE\left[e^{it \left(\sum_{i} f(Z_i, Z_{n+1}, Y_i)\right)}\right],
\end{align*}
where the second step follows from \eqref{eqn: O1}. Thus, we complete the proof.
\end{proof}

Note that the summands in $\sum_{i=1}^n f(Z_i,Z_{n+1},Y_i)$ remain dependent due to the presence of $Z_{n+1}$. However, the variance of this sum is straightforward to compute, which allows for the application of Chebyshev's inequality to obtain   polynomial concentration. In contrast,  directly dealing with the dependence in \eqref{eqn: o0}, where $J$ depends on $(Y_1,\dots,Y_n)$, is much more challenging.

\subsection{Proof Sketch of Theorem \ref{thm:local_non_convex}} 
We refer to Appendix \ref{app:one} for the detailed proof. Here, we provide only a proof sketch.

Our goal is to estimate  
$
    \operatorname{min}_{u\in\SS^{d-1}} u^{\top}\nabla^2 L(w) u,
$
where
\begin{equation}\label{eq:hessian}
    u^{\top}\nabla^2 L(w)u = \frac{1}{n}\sum_{i=1}^{n}(u^{\top}x_i)^2(3(w^{\top}x_i)^2-(w^{*\top}x_i)^2).
\end{equation}

The key observation is that as long as $w$ is not parallel to $w^{*}$, i.e. $\beta>0$, the random variable $3(w^{\top}x_i)^2-(w^{*\top}x_i)^2$ has an exponential tail on the negative side. Formally, we can prove that
\begin{equation}\label{eqn: x12}
\PP\left(3(w^{\top}x_i)^2-(w^{*\top}x_i)^2 \leq -t\right) \gtrsim e^{-C/\beta^2}\frac{\beta}{\sqrt{t}}e^{-Ct/\beta^2}
\end{equation}
for all $t \geq C$. The exponential tail in \eqref{eqn: x12} implies that 
\[
\min_{1\leq i\leq n}\left(3(w^{\top}x_i)^2-(w^{*\top}x_i)^2\right) = -\Theta(\beta^2\log n)
\]
holds with a high probability.

Let
$J:=\arg\min_{i\in [n]} \left(3(w^{\top}x_i)^2-(w^{*\top}x_i)^2\right)$ and choose $u_{J}=\frac{x_J}{\norm{x_J}{}}$. This yields
\begin{align*}
\min_{u\in\SS^{d-1}}u^{\top}\nabla^2 L(w)u &\leq u_{J}^{\top}\nabla^2 L(w)u_{J}\\ 
&=-\frac{1}{n}\norm{x_J}{}^2 \Theta(\beta^2\log n) + \frac{1}{n}\sum_{i\neq J}(u_J^{\top}x_i)^2\left(3(w^{\top}x_i)^2-(w^{*\top}x_i)^2\right).
\end{align*}
As $x_i\sim\cN(0,I_d)$, we should expect 
$\norm{x_J}{}^2=\Theta(d)$
with high probability. Furthermore, $\{x_i\}_{i\neq J}$ and $x_J$ should be ``nearly'' independent. Therefore, we should expect the following estimation 
\[
\frac{1}{n}\sum_{i\neq J}(u_J^{\top}x_i)^2\left(3(w^{\top}x_i)^2-(w^{*\top}x_i)^2\right)=\cO(1)
\]
with high probability. Combining all the estimation, we have
\[
\operatorname{min}_{u\in\SS^{d-1}} u^{\top}\nabla^2 L(w) u \leq -\frac{d}{n}\Theta(\beta^2\log n) +\cO(1)
\]
which is the desired result.

The key technical challenge lies in dealing with the dependence introduced by the adversarial index $J$. This can be handled by using the add-one trick introduced in Section \ref{sec: add-one}.

\subsection{Proof Sketch of Theorem \ref{thm:improved_local_non_convex}}
We refer to Appendix \ref{app:four} for the detailed proof. Here, we provide a sketch of that proof. 

Let  $\delta:=w-w^*$. Then we have
\begin{equation}\label{eqn: x23}
    u^{\top}\nabla^2 L(w)u = \frac{1}{n}\sum_{i=1}^{n}(u^{\top}x_i)^2(3(\delta^{\top}x_i)^2+6(\delta^{\top}x_i)(w^{*\top}x_i)+2(w^{*\top}x_i)^2).
\end{equation}
Analogous to the proof of Theorem \ref{thm:local_non_convex}, we shall adversarially choose a $\delta$ and a $u$ to make the above quantity as small as possible. To this end, we choose \[u_J=\frac{x_J}{\norm{x_J}{}} \quad\text{and}\quad
\delta_{J} = -\frac{x_J}{\norm{x_J}{}^2} w^{*\top}x_J\quad 
\] 
with $J:=\arg\max_{i\in [n]} w^{*\top}x_i$.
Plugging $u_J$ and $\delta_{J}$ into Eq.~\eqref{eqn: x23} gives 
\[
\begin{aligned}
    & \frac{1}{n}\sum_{i=1}^{n}(u_J^{\top}x_i)^2(3(\delta_J^{\top}x_i)^2+6(\delta_J^{\top}x_i)(w^{*\top}x_i)+2(w^{*\top}x_i)^2) \\
    &\quad\quad =-\frac{1}{n}\norm{x_J}{}^2 (w^{*\top}x_J)^2 + \frac{1}{n}\sum_{i\neq J}(u_J^{\top}x_i)^2(3(\delta_J^{\top}x_i)^2+6(\delta_J^{\top}x_i)(w^{*\top}x_i)+2(w^{*\top}x_i)^2).
\end{aligned}
\]
For the first term, we have $\|x_J\|=\Theta(\sqrt{d})$ and $w^{*\top}x_J= \Theta(\sqrt{\log n})$ with high probability, so the first term will be $-\Theta\left(\frac{d\log n}{n}\right)$. For the second term, as $\{x_i\}_{i\neq J}$ and $x_J$ are ``nearly'' independent, we have $\abs{\delta_J^{\top}x_i} =\widetilde{\cO}\left(\frac{1}{\sqrt{d}}\right)$ for $i\neq J$ and
\begin{equation}
 \frac{1}{n}\sum_{i\neq J}(u_J^{\top}x_i)^2(3(\delta_J^{\top}x_i)^2+6(\delta_J^{\top}x_i)(w^{*\top}x_i)+2(w^{*\top}x_i)^2)\approx \frac{1}{n}\sum_{i\neq J}\left(u_J^{\top}x_i\right)^2\left(w^{*\top}x_i\right)^2 = \cO(1)
\end{equation}
with high probability. Combining the two terms yields the desired result.

Again, the key technical challenge lies in dealing with the dependence introduced by the adversarial index $J$. This can be handled by the add-one trick introduced in Section \ref{sec: add-one}.

\subsection{Proof Sketch of Theorem \ref{thm: non one point convex}}
We refer to Appendix \ref{app:two} for the detailed proof. Here, we provide a sketch of that proof. 

Let  $\delta:=w-w^*$. Then, we have
\begin{equation}\label{eqn: x13}
\frac{\inn{\nabla L(w),w-w^*}}{\norm{w-w^*}{}^2}  =\frac{1}{n}\sum_{i=1}^{n}\frac{1}{\norm{\delta}{}^2}\left(\delta^{\top}x_i \right)^2({\delta^{\top}}x_i +2w^{*\top}x_i)({\delta^{\top}}x_i +w^{*\top}x_i) .
\end{equation}
Analogous to the proof of Theorem \ref{thm:local_non_convex}, we shall adversarially choose a $\delta$ to make the above quantity as small as possible. To this end, we choose 
\[
\delta_{J} = -\frac{3}{2}\frac{x_J}{\norm{x_J}{}^2} w^{*\top}x_J\quad \text{ with } J:=\arg\max_{i\in [n]} w^{*\top}x_i.
\] 
Plugging $\delta_{J}$ into Eq.~\eqref{eqn: x13} gives 
\[
\begin{aligned}
    &\frac{1}{n}\sum_{i=1}^{n}\frac{1}{\norm{\delta_{J}}{}^2}\left(\delta_J^{\top}x_i \right)^2({\delta_{J}^{\top}}x_i +2w^{*\top}x_i)({\delta_{J}^{\top}}x_i+w^{*\top}x_i) \\
    &\quad\quad =-\frac{1}{n}\Theta(1)\norm{x_J}{}^2 (w^{*\top}x_J)^2 + \frac{1}{n}\sum_{i\neq J}\frac{1}{\norm{\delta_{J}}{}^2}\left(\delta_{J}^{\top}x_i \right)^2({\delta_{J}^{\top}}x_i +2w^{*\top}x_i)({\delta_{J}^{\top}}x_i +w^{*\top}x_i).
\end{aligned}
\]
Similar to the proof of Theorem \ref{thm:improved_local_non_convex}, for the first term, we have $\|x_J\|=\Theta(\sqrt{d})$ and $w^{*\top}x_J= \Theta(\sqrt{\log n})$ with high probability, so the first term will be $-\Theta\left(\frac{d\log n}{n}\right)$. For the second term, leveraging the add-one trick, we have $\abs{\delta_J^{\top}x_i} =\widetilde{\cO}\left(\frac{1}{\sqrt{d}}\right)$ for $i\neq J$ and
\begin{equation}
\frac{1}{n}\sum_{i\neq J}\frac{1}{\norm{\delta_J}{}^2}\left(\delta_J^{\top}x_i \right)^2({\delta_J^{\top}}x_i +2w^{*\top}x_i)({\delta_J^{\top}}x_i +w^{*\top}x_i)\approx \frac{1}{n}\sum_{i\neq J}\frac{2}{\norm{\delta_J}{}^2}\left(\delta_J^{\top}x_i\right)^2\left(w^{*\top}x_i\right)^2 = \cO(1)
\end{equation}
with high probability. Combining the two terms yields the desired result.

\subsection{Proof Sketch of Theorem \ref{thm: one point convex positive}} 

The detailed proof is deferred to Appendix \ref{app:three}. Here we provide a sketch for it.

We expand the $\inn{\nabla L(w),w-w^*}$ term as
\begin{equation}\label{eq:onepointconvex}
    \inn{\nabla L(w),w-w^*} 
    =\frac{1}{n}\sum_{i=1}^{n}(\delta^{\top}x_i)^2(\delta^{\top}x_i +2w^{*\top}x_i)(\delta^{\top}x_i +w^{*\top}x_i)
\end{equation}
where we use $\delta$ to denote $w-w^*$. We want this quantity to be lower bounded.
Recall that in Theorem \ref{thm: non one point convex} the Gaussian tail of $w^{*\top}x_i$ causes the $\log n$ factor. It suggests that controlling the tail of $w^{*\top}x_i$ is crucial.

To this end, we divide $\{w^{*\top}x_i\}_{i=1}^n$ into two groups according to whether $w^{*\top}x_i$ is smaller or larger than a threshold  $t>0$. Let 
\[
\cI_{\leq}:=\left\{i\in [n]: \abs{w^{*\top}x_i} \leq t\right\}, \quad \cI_{\geq}:=\left\{i\in [n]: \abs{w^{*\top}x_i} > t\right\}. 
\]
Then the summation in Eq.~\eqref{eq:onepointconvex} can be partitioned into two groups.

\paragraph*{Step I.}
 We first lower bound
\[
\frac{1}{n}\sum_{i\in \cI_{\leq}}(\delta^{\top}x_i)^2(\delta^{\top}x_i +2w^{*\top}x_i)(\delta^{\top}x_i +w^{*\top}x_i)
\]
uniformly for all $\delta$. This is easier to deal with as the random summands are bounded for $i\in \cI_{\leq}$.

We do the following expansion
\begin{equation}\label{eqn: step1 expansion}
\begin{aligned}
&\frac{1}{n}\sum_{i\in \cI_{\leq}}(\delta^{\top}x_i)^2(\delta^{\top}x_i +2w^{*\top}x_i)(\delta^{\top}x_i +w^{*\top}x_i)\\
& \quad\quad = \frac{1}{n}\sum_{i\in \cI_{\leq}} (\delta^{\top}x_i)^4 + \frac{3}{n}\sum_{i\in \cI_{\leq}} (\delta^{\top}x_i)^3(w^{*\top}x_i) + \frac{2}{n}\sum_{i\in \cI_{\leq}} (\delta^{\top}x_i)^2 (w^{*\top}x_i)^2. 
\end{aligned}
\end{equation}
 Notice that the only term that can be negative is the second one. Our core idea is to use the first term and the third term to control the second term. To achieve this, we choose some constant $N>0$ and write
\[
\begin{aligned}
&\frac{3}{n}\sum_{i\in \cI_{\leq}} (\delta^{\top}x_i)^3(w^{*\top}x_i) = \frac{3}{n}\sum_{i\in \cI_{\leq}} (\delta^{\top}x_i)^31_{\abs{\Bar{\delta}^{\top}x_i}\leq N}(w^{*\top}x_i) + \frac{3}{n}\sum_{i\in \cI_{\leq}} (\delta^{\top}x_i)^31_{\abs{\Bar{\delta}^{\top}x_i}\geq N}(w^{*\top}x_i)\\
\end{aligned}
\]
where $\Bar{\delta}=\frac{\delta}{\norm{\delta}{}}$. Furthermore, we can lower bound the second term by
\[
\begin{aligned}
&\frac{3}{n}\sum_{i\in \cI_{\leq}} (\delta^{\top}x_i)^31_{\abs{\Bar{\delta}^{\top}x_i}\geq N}(w^{*\top}x_i)\\
&\quad\quad\quad\quad \geq -3\sqrt{\frac{1}{n}\sum_{i\in \cI_{\leq}}(\delta^{\top}x_i)^4}\sqrt{\frac{1}{n}\sum_{i\in \cI_{\leq}}(\delta^{\top}x_i)^21_{|\Bar{\delta}^{\top}x_i|\geq N}(w^{*\top}x_i)^2}\\
&\quad\quad\quad\quad \geq -\frac{1}{n}\sum_{i\in \cI_{\leq}} (\delta^{\top}x_i)^4-\frac{9}{4n}\sum_{i\in \cI_{\leq}}(\delta^{\top}x_i)^21_{|\Bar{\delta}^{\top}x_i|\geq N}(w^{*\top}x_i)^2 .
\end{aligned}
\]
Plug that in Eq.~\eqref{eqn: step1 expansion}, and we have
\[
\begin{aligned}
&\frac{1}{n}\sum_{i\in \cI_{\leq}}(\delta^{\top}x_i)^2(\delta^{\top}x_i +2w^{*\top}x_i)(\delta^{\top}x_i +w^{*\top}x_i) \\
&\quad\quad\geq \frac{3}{n}\sum_{i\in \cI_{\leq}} (\delta^{\top}x_i)^31_{\abs{\Bar{\delta}^{\top}x_i}\leq N}(w^{*\top}x_i)-\frac{9}{4n}\sum_{i\in \cI_{\leq}}(\delta^{\top}x_i)^21_{|\Bar{\delta}^{\top}x_i|\geq N}(w^{*\top}x_i)^2+ \frac{2}{n}\sum_{i\in \cI_{\leq}} (\delta^{\top}x_i)^2 (w^{*\top}x_i)^2 
\end{aligned}
\]

Finally, we concentrate the following two terms
\[
\frac{3}{n}\sum_{i\in \cI_{\leq}} (\delta^{\top}x_i)^31_{\abs{\Bar{\delta}^{\top}x_i}\leq N}(w^{*\top}x_i) \quad\text{ and }\quad \frac{2}{n}\sum_{i\in \cI_{\leq}} (\delta^{\top}x_i)^2 (w^{*\top}x_i)^2 .
\]
Note that when we are able to concentrate $\frac{2}{n}\sum_{i\in \cI_{\leq}} (\delta^{\top}x_i)^2 (w^{*\top}x_i)^2 $, we become able to concentrate $\frac{2}{n}\sum_{i\in \cI_{\leq}} (\delta^{\top}x_i)^21_{\abs{\Bar{\delta}^{\top}x_i}\leq N} (w^{*\top}x_i)^2$, then the $\frac{9}{4n}\sum_{i\in \cI_{\leq}}(\delta^{\top}x_i)^21_{|\Bar{\delta}^{\top}x_i|\geq N}(w^{*\top}x_i)^2$ term becomes $o_N(1)\norm{\delta}{}^2$ and we can just pick up a large absolute constant $N$ to make this term small.

For the term $
\frac{3}{n}\sum_{i\in \cI_{\leq}} (\delta^{\top}x_i)^31_{\abs{\Bar{\delta}^{\top}x_i}\leq N}(w^{*\top}x_i),
$
since all the elements are bounded, we can just invoke standard Hoeffding's inequality to concentrate for each $\delta$, then do a union bound on $\delta$ via some $\ep$-net arguments. For the term $
\frac{2}{n}\sum_{i\in \cI_{\leq}} (\delta^{\top}x_i)^2 (w^{*\top}x_i)^2,
$
we utilize the approximate independence between $\delta^{\top}x_i$ and $w^{*\top}x_i$ for most $\delta$, first regard $w^{*\top}x_i$ as constants, do concentration for uniformly all $\delta$ via Bernstein inequality plus union bound argument, then bound the $\ell^2$ and $\ell^{\infty}$ norm of $\left(w^{*\top}x_1,\dots,w^{*\top}x_n\right)$ and plug this in our Bernstein inequality to get the final bound. A detailed calculation reveals that we only need $n\gtrsim dt^2$ to concentrate these terms, which is definitely better than $n \gtrsim d\log d$ when we set the truncation level $t=o(\sqrt{\log d})$.

After the concentration step, we have
\[
\frac{1}{n}\sum_{i\in \cI_{\leq}}(\delta^{\top}x_i)^2(\delta^{\top}x_i +2w^{*\top}x_i)(\delta^{\top}x_i +w^{*\top}x_i)\gtrsim \norm{\delta}{}^2
\]
uniformly for all $\delta$.

\paragraph*{Step II.}
What remains is to lower bound the summation for $i\in \cI_{\geq}$:
\[
\begin{aligned}
\frac{1}{n}\sum_{i\in \cI_{\geq}}(\delta^{\top}x_i)^2(\delta^{\top}x_i +2w^{*\top}x_i)(\delta^{\top}x_i +w^{*\top}x_i) &\gtrsim
-\frac{1}{n}\sum_{i=1}^{n}(w^{*\top} x_i)^41_{|w^{*\top} x_i|\geq t}\\
&\approx -\EE\left[ (w^{*\top}x)^41_{|w^{*\top}x|\geq t}\right] = -\Theta(t^3 e^{-\frac{t^2}{2}}) .
\end{aligned}
\]
Therefore, Eq.~\eqref{eq:onepointconvex} can be lower bounded by $c\|\delta\|^2$ as long as $\|\delta\|^2\gtrsim t^3e^{-\frac{t^2}{2}}$,
 and the theorem follows directly.

\section{Conclusion}
In this paper, we present a fine-grained local-landscape analysis for phase retrieval under the regime of limited samples. On the negative side, we show that when $n=o(d\log d)$, the local landscape is neither convex nor one-point convex. Notably, the degree of non-convexity, measured by the smallest eigenvalue of Hessian matrix, becomes more pronounced as the dimension $d$ grows. On the positive side, we establish that as long as $n=\omega(d)$, the local landscape is one-point strongly convex outside a ball of radius $\cO\left(\sqrt{\log n/d}\right)$. When combined with prior results, this suggests that we can expect a provable global convergence for plain gradient descent with random initialization on Problem \eqref{eq:loss}, with the error being up to 
$o_d(1)$, in scenarios where $n=\omega(d)$.

For future works, it would be interesting to consider alternative properties that can guarantee local convergence, such as the Polyak-Lojasiewicz/Kurtyak-Lojasiewicz conditions~\citep{POLYAK,Kurdyka1998On}. It is also important to further explore   how our local landscape results can aid  in analyzing the entire dynamics of gradient descent. In addition, it would also be interesting to extend our analysis to the complex phase retrieval setting~\citep{candes2015}.

\paragraph*{Acknowledgements.} 
Lei Wu is supported in part by the National Key R\&D Program of China (No 2022YFA1008200).
Kaizhao Liu and Zihao Wang are partially supported by the elite undergraduate training program of School of Mathematical Sciences at Peking University. We thank Ziheng Cheng for helpful discussions and anonymous reviewers for constructive suggestions.


\bibliography{ref}

\newpage
\appendix
\clearpage
\appendix
\begin{center}
    \noindent\rule{\textwidth}{4pt} 
    \vspace{-0.2cm}
    \LARGE \textbf{Appendix} 
    \vspace{0.0cm}
    \noindent\rule{\textwidth}{1.2pt}
\end{center}

\startcontents[sections]
\printcontents[sections]{l}{1}{\setcounter{tocdepth}{2}}

\section{Proof of Theorem \ref{thm:local_non_convex}}\label{app:one}

We first recall the theorem.
\mainthmone*

Recalling Eq.~\eqref{eq:hessian}, we have

\begin{equation}\label{eqn: A1}
    u^{\top}\nabla^2 L(w)u = \frac{1}{n}\sum_{i=1}^{n}(u^{\top}x_i)^2(3(w^{\top}x_i)^2-(w^{*\top}x_i)^2):=\frac{1}{n}\sum_{i=1}^{n}(u^{\top}x_i)^2 z_i,
\end{equation}
where $x_i\stackrel{iid}{\sim} \cN(0,I_d)$ and we let $z_i:=3(w^{\top}x_i)^2-(w^{*\top}x_i)^2$. Note that $z_i$ is a quadratic function of two Gaussian random variables. 

The proof of Theorem \ref{thm:local_non_convex}  needs the following  lemma, whose proof is deferred to Appendix \ref{app A 1}. 

\begin{restatable}{lemma}{adversialkey}\label{lemma: nonconvex adversial choose}
Suppose that $\{(Z_i,Y_i)\}_{i=1}^n$ are \iid random variables with $Z_i\sim \cN(0,I_d)$, and there exist positive constants $C_1,C_2,C_3,C_4$ such that the distribution of $Y_i$ satisfies that for all $t\geq C_1$ and $\beta>0$,
\begin{equation}\label{eqn: A2}
\PP\left(Y_i \leq -t\right) \geq C_2e^{-C_3/\beta^2}\frac{\beta}{\sqrt{t}}e^{-C_3t/\beta^2}
\end{equation}
 and $$\EE\left[Y_i^2\right] \leq C_4.$$
Then, when $n\geq e^{3C_1C_3/\beta^2}$,  \wp at least $1-Ce^{-cd}-1/n-e^{-\sqrt{3C_3}C_2e^{-C_3/\beta^2}\sqrt{n}}$, it holds that
    \[\min_{u\in \SS^{d-1}} \frac{1}{n}\sum_{i=1}^n(u^{\top}Z_i)^2Y_i\leq -\frac{1}{24C_3}\beta^2\frac{d\log n }{n} +(1+\sqrt{3})\sqrt{C_4}.
    \]
\end{restatable}

\begin{remark}
Here, we explicitly state the dependence on the constants $C_1,C_2,C_3,C_4$ to clarify the influence of the tail behavior of $Y_i$. Particularly, the requirement $n\geq e^{3C_1C_3/\beta^2}$ arises because \eqref{eqn: A2} holds only when $t\geq C_1$.
\end{remark}


\paragraph*{Proof of Theorem \ref{thm:local_non_convex}.} By Lemma \ref{lemma: nonconvex adversial choose}, the proof follows the following two steps. 
\begin{itemize}
\item \textbf{\underline{Step I}.}
We first estimate the tail of the random variable $z_i$ on the negative side. Consider the decomposition $w=\alpha w^* + \beta w^{\perp}$ where $w^{\perp}$ is orthogonal to $w^*$ and normalized, i.e., $\|w^{\perp}\|=1$. Let
$W_i^{\perp}:=(w^\perp)^{\top}x_i$ and $W_i^*:=w^{*\top}x_i$. Then, $W_i^\perp$ and $W_i^*$ are two independent $\cN(0,1)$ random variables and moreover,
\[
z_i=3(w^{\top}x_i)^2-(w^{*\top}x_i)^2= 3\beta^2W_i^{\perp 2}+6\alpha\beta W_i^{\perp}W_i^*+(3\alpha^2-1)W_i^{*2}:=f(W_i^{\perp}, W_i^*).
\]
Since $w\in\rloc$, as defined in Eq.~\eqref{eqn: local}, it follows that  $\alpha\in (2/3,4/3)$ and $0<\beta\leq 1$. 

Note that  $z_i=f(W_i^{\perp}, W_i^*)$ is a quadratic form of $(W_i^{\perp},W_i^*)$, which satisfies the assumptions of Lemma \ref{lemma: lower exp tail for quadratic form}. By using the same notation in Lemma \ref{lemma: lower exp tail for quadratic form}, we have 
\begin{equation}\label{eqn: A3}
0<\lambda_+\leq C,\qquad \lambda_-=-\frac{6\beta^2}{\left(3\beta^2+3\alpha^2-1+\sqrt{(3\beta^2-3\alpha^2+1)^2 + 36\beta^2}\right)} \lesssim -\beta^2,
\end{equation}
where the last inequality uses the assumption that $\alpha\in (2/3,4/3)$ and $0<\beta\leq 1$.

Applying Lemma \ref{lemma: lower exp tail for quadratic form}, we have for all $t\geq C$, it holds that
\begin{equation}\label{eqn: neg-1}
\PP\left(z_i \leq -t\right) \gtrsim e^{-C/\beta^2}\frac{\beta}{\sqrt{t}}e^{-Ct/\beta^2 } .
\end{equation}
In addition, it is easy to check that $\EE[z_i^2]\leq C$ for all $i\in [n]$ due to  the boundedness assumption on $\alpha$ and $\beta$. Therefore, $z_i$'s satisfy the condition on $Y_i$'s in Lemma \ref{lemma: nonconvex adversial choose}.

\item \textbf{\underline{Step II}.} Let $Q\in\RR^{d\times (d-2)}$ be an orthonormal basis of $\spg{w,w^*}^\perp$. Then, by Eq.~\eqref{eqn: A1} we have that 
\begin{align*}
    \min_{u\in\SS^{d-1}} u^{\top}\nabla^2 L(w)u&\leq \min_{u\in\SS^{d-1}\cap \spg{w,w^*}^\perp} u^{\top}\nabla^2 L(w)u \\ 
    &=\min_{v\in\SS^{d-3}} \frac{1}{n}\sum_{i=1}^{n}(v^{\top} \tilde{x}_i)^2 z_i \quad\quad \qquad (\text{let }u=Qv, \tilde{x}_i=Q^\top x_i)\\ 
    &\leq -c\beta^2\frac{(d-2)\log n}{n}+C, \qquad  (\text{by Lemma \ref{lemma: nonconvex adversial choose}})
\end{align*}
holds \wp at least $1-Ce^{-cd}-C/n-e^{-ce^{-C/\beta^2}\sqrt{n}}$. Note that the first step ensures $\tilde{x}_i:=Q^{\top}x_i$ are independent from $z_i=3(w^{\top}x_i)^2-(w^{*\top}x_i)^2$ and therefore, we can apply Lemma \ref{lemma: nonconvex adversial choose}.
\qed

\end{itemize}

\subsection{Proof of Lemma \ref{lemma: nonconvex adversial choose}}
\label{app A 1}

\paragraph*{The add-one trick.}
Define the random index: 
\begin{equation}\label{eqn: A4}
J=\arg\min_{i\in [n]}Y_i. 
\end{equation}
It is obvious that $J$ is independent of $\{Z_i\}_{i=1}^n$ as  $J\in \sigma(Y_1,\cdots,Y_n)$. Taking $u=\frac{Z_J}{\norm{Z_J}{}}$ gives 
\begin{align}
\notag    \min_{u\in \SS^{d-1}} \frac{1}{n}\sum_{i=1}^n(Z_i^{\top}u)^2Y_i&\leq \frac{1}{n}\sum_{i=1}^n\left(Z_i^{\top}\frac{Z_J}{\|Z_J\|}\right)^2Y_i  \\ 
\notag &=\underbrace{\frac{1}{n}\left(\|Z_J\|^2-\left(Z_{n+1}^{\top}\frac{Z_J}{\|Z_J\|}\right)^2\right)Y_J}_{I_1}\\ 
        &\qquad\qquad +\underbrace{\frac{1}{n}\left(\sum_{i\ne J} \left(Z_i^{\top}\frac{Z_J}{\|Z_J\|}\right)^2Y_i+\left(Z_{n+1}^{\top}\frac{Z_J}{\|Z_J\|}\right)^2Y_J\right)}_{I_2},
\end{align}
where the second step adopts the add-one decomposition \eqref{eqn: add-one}.
        

\paragraph*{\underline{Bound $I_1$}.}
By Lemma \ref{lemma: same-distribution}, we have $Z_J\overset{d}{=}Z_i\sim \cN(0,I_d)$. Applying  Lemma \ref{lemma: concetration of the norm}, we  \wp at least $1-Ce^{-cd}$ that  
\begin{equation}\label{eqn: A66}
\|Z_J\|^2 \geq \frac{1}{4}d.
\end{equation}
Since $Z_J$ is independent of $Z_{n+1}$,  we have $Z_{n+1}^{\top}\frac{Z_J}{\norm{Z_J}{}}\sim \cN(0,1)$. Applying Lemma \ref{lemma: tail of gaussian}, \wp at least $1-Ce^{-cd}$, we have
\begin{equation}\label{eqn: A7}
\left(Z_{n+1}^{\top}\frac{Z_J}{\norm{Z_J}{}}\right)^2 \leq \frac{1}{8}d.
\end{equation}

By the assumption that for $t\geq C_1$, $
\PP\left(Y_i \leq -t\right) \geq C_2 e^{-C_3/\beta^2}\frac{\beta}{\sqrt{t}}e^{-C_3t/\beta^2}$, we have 
    \begin{align}\label{eqn: 444}
    \notag    \PP(Y_J \geq -t)&\leq (1-\PP(Y\leq -t))^n \\
    \notag        &\leq \left(1-C_2e^{-C_3/\beta^2}\frac{\beta}{\sqrt{t}}e^{-C_3t/\beta^2}\right)^n\\
    \notag        &= e^{n\log \left(1-C_2e^{-C_3/\beta^2}\frac{\beta}{\sqrt{t}}e^{-Ct/\beta^2}\right)}\\
                  &\leq e^{-C_2e^{-C_3/\beta^2}\frac{\beta}{\sqrt{t}}e^{-C_3t/\beta^2}n},
    \end{align}
where the first step uses the independence among $Y_1,\dots,Y_n$.

Next we shall take $t$ such that $\frac{\beta}{\sqrt{t}}e^{-C_3t/\beta^2}\gtrsim \frac{1}{\sqrt{n}}$. Specifically, consider $t=q\beta^2\log n$, where $q$ is a positive constant to be determined latter. Then, we have
\[
\frac{\beta}{\sqrt{t}}e^{-C_3t/\beta^2} = \sqrt{\frac{1}{q\log n}} e^{-qC_3\log n} = \sqrt{\frac{1}{q\log n}} \frac{1}{n^{qC_3}}
\]
and thus, we can take $q=1/(3C_3)$ such that 
\begin{equation}\label{eqn: yy1}
\frac{\beta}{\sqrt{t}}e^{-C_3t/\beta^2}\geq \sqrt{3C_3} \frac{1}{\sqrt{\log n}n^{1/3}}\geq \sqrt{\frac{3C_3}{n}}.
\end{equation}
Given the condition $t=q\beta^2\log n\geq C_1$, the above estimate requires $n\geq e^{3C_1C_3/\beta^2}$.

Plugging \eqref{eqn: yy1} into \eqref{eqn: 444} gives that when $n\geq e^{3C_1C_3/\beta^2}$,
\begin{equation}\label{eqn: A8}
\PP\left(Y_J\geq -\frac{\beta^2\log n}{3C_3}\right) \leq e^{-\sqrt{3C_3}C_2e^{-C_3/\beta^2}\sqrt{n}} .
\end{equation}


Combining \eqref{eqn: A66}, \eqref{eqn: A7} and \eqref{eqn: A8}, we have if $n\geq e^{3C_1C_3/\beta^2}$, \wp at least $1-Ce^{-cd}-e^{-\sqrt{3C_3}C_2e^{-C_3/\beta^2}\sqrt{n}}$, 
\begin{equation}\label{eqn: A12}
I_1=\frac{1}{n}\left(\|Z_J\|^2-\left(Z_{n+1}^{\top}\frac{Z_J}{\|Z_J\|}\right)^2\right)Y_J=\frac{1}{n}\left(\frac{d}{4}-\frac{d}{8}\right)(-t) \leq -\frac{1}{24C_3}\beta^2\frac{d\log n}{n} .
\end{equation}

\paragraph*{\underline{Bound $I_2$}.}
    By Lemma \ref{lemma: add-one}, 
    \begin{equation}\label{eqn: A9}
    I_2
    \overset{d}{=} \frac{1}{n}\sum_{i=1}^n \left(Z_i^{\top}\frac{Z_{n+1}}{\|Z_{n+1}\|}\right)^2Y_i.
    \end{equation}

For simplicity, let $U_i=Z_i^{\top}Z_{n+1}/\|Z_{n+1}\|$. By Lemma \ref{lem: gaussian inner product}, $U_i\stackrel{iid}{\sim} \cN(0,1)$ for $i=1,2,\dots,n$.
Noting that $(Y_1,\dots,Y_n)$ and $(U_1,\dots,U_{n+1})$ are independent and $Y_1,\dots,Y_n$ are \iid random variable, we have
\begin{align*}
\EE[I_2] &= \EE\left[\frac 1 n\sum_i U_i^2 Y_i\right] = \frac 1 n\sum_i\EE[ U_i^2] \EE[Y_i]=\EE[Y_1],\\ 
\var[I_2] &= \var\left[\frac 1 n\sum_i U_i^2 Y_i\right] = \frac{1}{n^2}\sum_i \var[U_i^2Y_i] + \frac{1}{n^2}\sum_{i\neq j}\cov\left(U_i^2Y_i,U_j^2Y_j\right) \\ 
 &\stackrel{(i)}{=}  \frac{1}{n^2}\sum_i \var[U_i^2Y_i] \\ 
&= \frac{1}{n} \var[U_1^2Y_1] \leq\frac{1}{n} \EE[U_1^4Y_1^2]= \frac{1}{n} \EE[U_1^4]\EE[Y_1^2]\stackrel{(ii)}{\leq}\frac{3C_4}{n},
\end{align*}
where $(i)$ follows from the fact that for $i\neq j$,
\[
    \cov\left(U_i^2Y_i,U_j^2Y_j\right) = \EE[U_i^2Y_i U_j^2Y_j] - \EE[U_i^2Y_i]\EE[U_j^2Y_j] = 0
\]
due to the independence between $U_i^2 Y_i$ and $U_j^2Y_j$, and $(ii)$ uses the assumption $\EE\left[Y^2_i\right]\leq C_4$.


Then, by Chebyshev's inequality, we have 
\[
\PP\left(\abs{I_2-\EE[Y_1]}\geq \lambda\right)\leq \frac{\var[I_2]}{\lambda^2}\lesssim \frac{3C_4}{n\lambda^2}.
\]
By letting $\lambda=\sqrt{3C_4}$,   we have \wp at least $1-1/n$, it holds that
\begin{equation}\label{eqn: A13}
I_2\leq \EE[Y_1] + \sqrt{3C_4} \leq\sqrt{\EE[Y_1^2]}+\sqrt{3C_4}\leq (1+\sqrt{3})\sqrt{C_4}.
\end{equation}

Combining \eqref{eqn: A12} and \eqref{eqn: A13}, we complete the proof.



\qed

\section{Proof of Theorem \ref{thm:improved_local_non_convex}}\label{app:four}
We first recall the theorem.
\mainthmfour*
\begin{proof}
    Denote $\delta:=w-w^*$. By a simple calculation, we have
    \begin{equation}\label{equation: x23}
   u^{\top}\nabla^2 L(w)u = \frac{1}{n}\sum_{i=1}^{n}(u^{\top}x_i)^2(3(\delta^{\top}x_i)^2+6(\delta^{\top}x_i)(w^{*\top}x_i)+2(w^{*\top}x_i)^2).
    \end{equation}
    Our subsequent estimates are based on the following relaxation:
    \begin{equation}\label{eqn: E1}
    \min_{u\in\SS^{d-1},\|\delta\|\leq \gamma_{n,d}} u^{\top}\nabla^2 L(w) u\leq \min_{\|\delta\|\leq \gamma_{n,d}, \langle \delta,w^*\rangle=0}  \frac{\delta^{\top}\nabla^2 L(w)\delta}{\norm{\delta}{}^2},
    \end{equation}
    for the latter, $\delta^{\top}x_i$ and $w^{*\top}x_i$ become independent random variables as $\<\delta, w^{*}\>=0$.
    
    Next, we shall use the following lemma to bound the RHS in Eq.~\eqref{eqn: E1}, whose  proof is deferred to Appendix \ref{app E 1}.

\begin{restatable}{lemma}{lemmaimprovednoconvex}\label{lemma: improved no convex}
    Let $\{(Z_i,Y_i)\}_{i=1}^n$ be \iid random variables with $Z_i\sim \cN(0,I_d)$ and $Y_i\sim \cN(0,1)$. If $C\sqrt{\frac{\log n}{d}}\leq \gamma \leq 1$, then w.p. at least $1-Ce^{-c\sqrt{d}}-Ce^{-c\log n}$, we have 
    \[\min_{\|\delta\|\leq \gamma} \frac{1}{n}\sum_{i=1}^n\left(Z_i^{\top}\frac{\delta}{\norm{\delta}{}}\right)^2(3(Z_i^{\top}\delta)^2+6(Z_i^{\top}\delta)Y_i +2Y_i^2)\leq -c\frac{d\log n}{n}+C .\]
\end{restatable}

Then, by \eqref{equation: x23}, \eqref{eqn: E1}, and Lemma \ref{lemma: improved no convex}, we have  when $n,d\geq C$, 
\[
\min_{u\in\SS^{d-1},\|\delta\|\leq \gamma_{n,d}} u^{\top}\nabla^2 L(w) u\leq \min_{\|\delta\|\leq \gamma_{n,d}, \langle \delta,w^*\rangle=0}  \frac{\delta^{\top}\nabla^2 L(w)\delta}{\norm{\delta}{}^2} \leq -c\frac{d\log n}{n}+C 
\]
\wp at least $1-Ce^{-c\sqrt{d}}-Ce^{-c\log n}$. This completes the proof.
\end{proof}

\subsection{Proof of Lemma \ref{lemma: improved no convex}}\label{app E 1}

\paragraph*{\underline{Step 1}: Choose an adversarial direction.}
Let $J:=\arg\max_{i\in [n]}Y_i$. As $J\in \sigma(Y_1,\cdots,Y_n)$, $J$ is independent of $Z_1,\dots,Z_n$. Let
\begin{equation}\label{eqn: gamma01}
\delta_0 =-\gamma_0 \frac{Z_J}{\|Z_J\|}, \text{ with } \gamma_0=\frac{\abs{\EE\left[Y_J\right]}}{\sqrt{d}}.
\end{equation}
Since $\EE[Y_J]=\EE[\max(Y_1,\dots,Y_n)]=\Theta(\sqrt{\log n})$ (see, e.g.,~\citet[Exercise 2.5.10 and 2.5.11]{vershynin2018high}), we have 
\[
\|\delta_0\|=\gamma_0\leq \gamma\leq 1.
\]

Thus, we can take $\delta=\delta_0$ and consequently,
\begin{align*}
       \min_{\|\delta\|\leq \gamma}\frac{1}{n}\sum_{i=1}^n&\left(Z_i^{\top}\frac{{\delta}}{\norm{\delta}{}}\right)^2(3(Z_i^{\top}\delta)^2+6(Z_i^{\top}\delta)Y_i +2Y_i^2)\\
       &\leq \frac{1}{n}\sum_{i=1}^n\left(Z_i^{\top}\frac{Z_J}{\|Z_J\|}\right)^2\left(3(\gamma_0Z_i^{\top}\frac{Z_J}{\|Z_J\|})^2-6(\gamma_0Z_i^{\top}\frac{Z_J}{\|Z_J\|})Y_i +2Y_i^2 \right)\\ 
       &= \frac{1}{n}\sum_{i=1}^n f(Z_i,Z_J, Y_i),
  \end{align*}
  where 
\[
  f(z,z',y) = \left(z^{\top}\frac{z'}{\|z'\|}\right)^2 \left(3\left(\gamma_0 z^{\top}\frac{z'}{\|z'\|}\right)^2 -6\gamma_0 z^{\top}\frac{z'}{\|z'\|}y+2y^2\right).
\]

Next, we shall follow the add-one trick in Section \ref{sec: add-one} to bound $I_1$ and $I_2$ separately.

\subsubsection*{\underline{Step 2: Bound $I_1$}.}  
By Lemma \ref{lemma: same-distribution},  $Z_J\overset{d}{=}Z_i\sim \cN(0,I_d)$ and is independent of $Z_{n+1}$. Lemma \ref{lemma: concetration of the norm} implies \wp at least $1-Ce^{-c\sqrt{d}}$, it holds that 
\begin{equation}\label{eqn: B6}
\sqrt{d}-\sqrt[4]{d} \leq \|Z_J\|\leq \sqrt{d}+\sqrt[4]{d}. 
\end{equation}
Lemma \ref{lemma: gaussian extreme concentration} implies \wp at least $1-2e^{-c\log n}$, we have
\begin{equation}\label{eqn: E4}
\frac{7}{8}\EE\left[Y_J\right] \leq Y_J \leq \frac{9}{8}\EE\left[Y_J\right].
\end{equation}
Thus, combining these estimates, we have when $d\geq C$ \wp at least $1-Ce^{-c\sqrt{d}}-2e^{-c\log n}$
\begin{align}\label{eqn: E5}
f(Z_J,Z_J, Y_J) &= \|Z_J\|^2\left(3\gamma_0^2\|Z_J\|^2 - 6\gamma_0 \|Z_J\| Y_J+2Y^2_J\right) \notag\\ 
&\leq \|Z_J\|^2 \left(\frac{3\|Z_J\|^2}{d} - \frac{6\|Z_J\|}{\sqrt{d}}\cdot \frac{7}{8} + \frac{81}{64}\right) \left(\EE[Y_J]\right)^2\notag\\
&\lesssim - d\log n.
\end{align}

Let $Q=Z_{n+1}^{\top}\frac{Z_J}{\norm{Z_J}{}}\sim \cN(0,1)$. By Lemma \ref{lemma: tail of gaussian}, we have
 \wp at least $1-Ce^{-c\sqrt{d}}$,  
\begin{equation}\label{eqn: E10}
\abs{Q} \leq \sqrt[4]{d}.
\end{equation}
 In that case, we have when $d\geq C$, 
\begin{align}\label{eqn: E7}
\notag f(Z_{n+1}, Z_J, Y_J)&=Q^2\left(3\gamma_0^2Q^2 - 6\gamma_0 Q Y_J+2Y^2_J\right)\\
& \stackrel{\text{ use \eqref{eqn: gamma01}}}{=} Q^2\left(3\frac{(\EE\left[Y_J\right])^2}{d}Q^2 - 6\frac{\abs{\EE\left[Y_J\right]}}{\sqrt{d}} Q Y_J+2Y^2_J\right) \notag \\ 
&\stackrel{\text{use \eqref{eqn: E4} and \eqref{eqn: E10}}}{\geq }Q^2 \left(-\frac{27}{4\sqrt[4]{d}}+\frac{98}{64}\right)\EE[Y_J]^2 \notag \\
&\geq 0.
\end{align}
Combing \eqref{eqn: E5} and \eqref{eqn: E7}, we have 
\begin{align}\label{eqn: E-I_1}
I_1=\frac{1}{n}\left(f(Z_J,Z_J, Y_J)-f(Z_{n+1}, Z_J, Y_J)\right)&\leq -\frac{1}{n}\left( c\log n \|Z_J\|^2 + 0\right)\leq -c \frac{d\log n}{n} .
\end{align}

\subsubsection*{\underline{Step 3: Bound $I_2$.}}

For simplicity, let $U_i := Z_i^{\top}\frac{Z_{n+1}}{\norm{Z_{n+1}}{}}$. By Lemma \ref{lem: gaussian inner product}, we have $U_i\stackrel{iid}{\sim}\cN(0,1)$ for $i=1,\dots,n$.
By Lemma \ref{lemma: add-one}, 
\begin{align*}
I_2 \overset{d}{=}\frac{1}{n}\sum_{i=1}^n f(Z_i,Z_{n+1}, Y_i)& = \frac{1}{n}\sum_{i=1}^n U_i^2(3\gamma_0^2U_i^2-6\gamma_0U_iY_i+2Y_i^2)\\ 
  &\overset{d}{=}\frac{1}{n}\sum_{i=1}^n U_i^2(3\gamma_0^2U_i^2+6\gamma_0U_iY_i+2Y_i^2),
\end{align*}
where the last step uses the fact that $(Y_1,\dots,Y_n)$ are independent of $(U_1,\dots,U_n)$ and that the distribution of $Y_i$ is symmetric around zero.

Since $U_i\stackrel{iid}{\sim}\cN(0,1)$, $Y_i\stackrel{iid}{\sim}\cN(0,1)$, and $\gamma_0\leq 1$,
we have
\[
    \EE[I_2] = \EE\left[\frac{1}{n}\sum_{i=1}^n U_i^2(3\gamma_0^2U_i^2+6\gamma_0U_iY_i+2Y_i^2)\right]\leq C
\]
and 
  \[
    \begin{aligned}
    \var[I_2]&=\operatorname{Var}\left(\frac{1}{n}\sum_{i=1}^n U_i^2(3\gamma_0^2U_i^2+6\gamma_0U_iY_i+2Y_i^2)\right)\\
    &  =\frac{1}{n^2}\sum_{i=1}^n \operatorname{Var}\left(U_i^2(3\gamma_0^2U_i^2+6\gamma_0U_iY_i+2Y_i^2)\right)\\ 
    & \quad + \frac{1}{n^2}\sum_{i\neq j}\operatorname{Cov}\left(U_i^2(3\gamma_0^2U_i^2+6\gamma_0U_iY_i+2Y_i^2),U_j^2(3\gamma_0^2U_j^2+6\gamma_0U_jY_j+2Y_j^2)\right)\\
     &  \stackrel{(i)}{=}\frac{1}{n^2}\sum_{i=1}^n \operatorname{Var}\left(U_i^2(3\gamma_0^2U_i^2+6\gamma_0U_iY_i+2Y_i^2)\right)+ 0\\
     &   =\frac{1}{n}\operatorname{Var}\left(U_1^2(3\gamma_0^2U_1^2+6\gamma_0U_1Y_1+2Y_1^2)\right)\\
    &  \leq \frac{C}{n},
    \end{aligned}
    \]
where $(i)$ follows from the independence between $U_i^2(3\gamma_0^2U_i^2+6\gamma_0U_iY_i+2Y_i^2)$ and $U_j^2(3\gamma_0^2U_j^2+6\gamma_0U_jY_j+2Y_j^2)$ for $i\neq j$.

By Chebyshev's inequality, we have
\begin{equation}\label{eqn: E-I2}
\PP\left(|I_2-\EE[I_2]|\geq t\right) \leq \frac{\var[I_2]}{t^2}\leq \frac{C}{nt^2}.
\end{equation}
Therefore, $I_2 \leq C$ \wp at least $1-C/n$.

Combining \eqref{eqn: E-I_1} and \eqref{eqn: E-I2}, we have \wp 
 at least $1-Ce^{-c\sqrt{d}}-Ce^{-c\log n}$ that
\[
I_1+I_2\leq -c \frac{d\log n}{n} + C.
\]
\qed

\section{Proof of Theorem \ref{thm: non one point convex}}\label{app:two}
The proof of Theorem \ref{thm: non one point convex} is similar to the proof of Theorem \ref{thm:improved_local_non_convex}. We first recall the theorem.
\mainthmtwo*
\begin{proof}
    Let $\delta=w-w^*$. By a simple calculation, we have
    \begin{equation}\label{equation: one point convexity}
    \frac{\inn{\nabla L(w),w-w^*}}{\norm{w-w^*}{}^2}  =\frac{1}{n}\sum_{i=1}^{n}\frac{1}{\norm{\delta}{}^2}\left(\delta^{\top}x_i \right)^2({\delta^{\top}}x_i +2w^{*\top}x_i)({\delta^{\top}}x_i +w^{*\top}x_i) .
    \end{equation}
    Our subsequent estimates are based on the following relaxation:
    \begin{equation}\label{eqn: B1}
    \min_{\|\delta\|\leq \gamma} \frac{\inn{\nabla L(w),w-w^*}}{\norm{w-w^*}{}^2}\leq \min_{\|\delta\|\leq \gamma, \langle \delta,w^*\rangle=0} \frac{\inn{\nabla L(w),w-w^*}}{\norm{w-w^*}{}^2},
    \end{equation}
    for the latter, $\delta^{\top}x_i$ and $w^{*\top}x_i$ become independent random variables as $\<\delta, w^{*}\>=0$.
    
    Next we shall use the following lemma to bound the RHS in Eq.~\eqref{eqn: B1}, whose  proof is deferred to Appendix \ref{app B 1}.

\begin{restatable}{lemma}{lemmanoonepointconvex}\label{lemma: no one point convex}
    Let $\{(Z_i,Y_i)\}_{i=1}^n$ be \iid random variables with $Z_i\sim \cN(0,I_d)$ and $Y_i\sim \cN(0,1)$. If $C\sqrt{\frac{\log n}{d}}\leq \gamma \leq 1$, then \wp at least $1-Ce^{-c\sqrt{d}}-Ce^{-c\log n}$, we have 
    \[\min_{\|\delta\|\leq \gamma} \frac{1}{n}\sum_{i=1}^n\left(Z_i^{\top}\frac{\delta}{\norm{\delta}{}}\right)^2(Z_i^{\top}\delta +2Y_i)(Z_i^{\top}\delta+Y_i)\leq -c\frac{d\log n}{n}+C .\]
\end{restatable}

Then, by \eqref{equation: one point convexity}, \eqref{eqn: B1},  and Lemma \ref{lemma: no one point convex}, we have  when $n,d\geq C$, 
\[
\min_{\|\delta\|\leq \gamma} \frac{\inn{\nabla L(w),w-w^*}}{\norm{w-w^*}{}^2}\leq \min_{\|\delta\|\leq \gamma, \langle \delta,w^*\rangle=0} \frac{\inn{\nabla L(w),w-w^*}}{\norm{w-w^*}{}^2} \leq -c\frac{d\log n}{n}+C
\]
\wp at least $1-Ce^{-c\sqrt{d}}-Ce^{-c\log n}$. This completes the proof.
\end{proof}

\subsection{Proof of Lemma \ref{lemma: no one point convex}}\label{app B 1}

\paragraph*{\underline{Step 1}: Choose an adversarial direction.}
Let $J:=\arg\max_{i\in [n]}Y_i$. As $J\in \sigma(Y_1,\cdots,Y_n)$, $J$ is independent of $Z_1,\dots,Z_n$. Let
\begin{equation}\label{eqn: gamma0}
\delta_0 =-\gamma_0 \frac{Z_J}{\|Z_J\|}, \text{ with } \gamma_0=\frac{3}{2}\frac{\abs{\EE\left[Y_J\right]}}{\sqrt{d}}.
\end{equation}
Since $\EE[Y_J]=\EE[\max(Y_1,\dots,Y_n)]=\Theta(\sqrt{\log n})$ (see, e.g.,~\citet[Exercise 2.5.10 and 2.5.11]{vershynin2018high}), we have 
\[
\|\delta_0\|=\gamma_0\leq \gamma\leq 1. 
\]
Thus, we can take $\delta=\delta_0$ and consequently,
\begin{align*}
       \min_{\|\delta\|\leq \gamma}\frac{1}{n}\sum_{i=1}^n&\left(Z_i^{\top}\frac{{\delta}}{\norm{\delta}{}}\right)^2(Z_i^{\top}\delta +2Y_i)(Z_i^{\top}\delta+Y_i)\\
       &\leq \frac{1}{n}\sum_{i=1}^n\left(Z_i^{\top}\frac{Z_J}{\|Z_J\|}\right)^2\left(-\gamma_0Z_i^{\top}\frac{Z_J}{\|Z_J\|} +2Y_i\right)\left(-\gamma_0 Z_i^{\top}\frac{Z_J}{\|Z_J\|}+Y_i\right)\\ 
       &= \frac{1}{n}\sum_{i=1}^n f(Z_i,Z_J, Y_i),
  \end{align*}
  where 
\[
  f(z,z',y) = \left(z^{\top}\frac{z'}{\|z'\|}\right)^2 \left(-\gamma_0 z^{\top}\frac{z'}{\|z'\|} + 2y\right)\left(-\gamma_0 z^{\top}\frac{z'}{\|z'\|}+y\right).
\]


Next, we shall follow the add-one trick in Appendix \ref{sec: add-one} to bound $I_1$ and $I_2$ separately.

\subsubsection*{\underline{Step 2: Bound $I_1$}.}  
By Lemma \ref{lemma: same-distribution},  $Z_J\overset{d}{=}Z_i\sim \cN(0,I_d)$ and is independent of $Z_{n+1}$. Lemma \ref{lemma: concetration of the norm} implies \wp at least $1-Ce^{-c\sqrt{d}}$, it holds that 
\begin{equation}
\sqrt{d}-\sqrt[4]{d} \leq \|Z_J\|\leq \sqrt{d}+\sqrt[4]{d}. 
\end{equation}
Lemma \ref{lemma: gaussian extreme concentration} implies \wp at least $1-2e^{-c\log n}$, we have
\begin{equation}\label{eqn: B4}
\frac{7}{8}\EE\left[Y_J\right] \leq Y_J \leq \frac{9}{8}\EE\left[Y_J\right].
\end{equation}
Thus, combing these estimates leads to \wp at least $1-Ce^{-c\sqrt{d}}-2e^{-c\log n}$, we have when $d\geq C$, 
\begin{align}\label{eqn: B5}
f(Z_J,Z_J, Y_J) &= \|Z_J\|^2\left(\gamma_0^2\|Z_J\|^2 - 3\gamma_0 \|Z_J\| Y_J+2Y^2_J\right) \notag\\ 
&\leq \|Z_J\|^2 \left(\frac{9\|Z_J\|^2}{4d} - \frac{9\|Z_J\|}{2\sqrt{d}}\cdot \frac{7}{8} + \frac{81}{64}\right) \EE[Y_J]^2\notag\\
&\lesssim - d\log n.
\end{align}

Let $Q=Z_{n+1}^{\top}\frac{Z_J}{\norm{Z_J}{}}\sim \cN(0,1)$. By Lemma \ref{lemma: tail of gaussian}, we have
 \wp at least $1-Ce^{-c\sqrt{d}}$,  
\begin{equation}\label{eqn: B10}
\abs{Q} \leq \sqrt[4]{d}.
\end{equation}
 In that case, we have when $d\geq C$, 
\begin{align}\label{eqn: B7}
\notag f(Z_{n+1}, Z_J, Y_J)&=Q^2\left(-\gamma_0 Q +2Y_J\right)\left(-\gamma_0 Q+Y_J\right)\\
& \stackrel{\text{ use \eqref{eqn: gamma0}}}{=} Q^2\left(-\frac{3Q}{2\sqrt{d}}\EE[Y_J]  +2Y_J\right)\left(- \frac{3Q}{2\sqrt{d}}\EE[Y_J]+Y_J\right) \notag \\ 
&\stackrel{\text{use \eqref{eqn: B4} and \eqref{eqn: B10}}}{\geq }Q^2 \left(-\frac{3}{2\sqrt[4]{d}} +\frac{7}{4}\right)\left(- \frac{3}{2\sqrt[4]{d}}+\frac{7}{8}\right)\EE[Y_J]^2 \notag \\
&\geq 0.
\end{align}
Combing \eqref{eqn: B5} and \eqref{eqn: B7}, we have 
\begin{align}\label{eqn: B-I_1}
I_1=\frac{1}{n}\left(f(Z_J,Z_J, Y_J)-f(Z_{n+1}, Z_J, Y_J)\right)&\leq -\frac{1}{n}\left( c\log n \|Z_J\|^2 + 0\right)\leq -c \frac{d\log n}{n} .
\end{align}

\subsubsection*{\underline{Step 3: Bound $I_2$.}}

Let $U_i = Z_i^{\top}\frac{Z_{n+1}}{\norm{Z_{n+1}}{}}$. By Lemma \ref{lem: gaussian inner product}, we have $U_i\stackrel{iid}{\sim}\cN(0,1)$ for $i=1,\dots,n$.

By Lemma \ref{lemma: add-one}, 
\begin{align*}
I_2 \overset{d}{=}\frac{1}{n}\sum_{i=1}^n f(Z_i,Z_{n+1}, Y_i)& = \frac{1}{n}\sum_{i=1}^n U_i^2(-\gamma_0U_i+2Y_i)(-\gamma_0 U_i+Y_i)\\ 
  &\overset{d}{=}\frac{1}{n}\sum_{i=1}^n U_i^2(\gamma_0U_i+2Y_i)(\gamma_0 U_i+Y_i),
\end{align*}
where the last step use the independence between $(Z_1,\dots,Z_n)$ and $(Y_1,\dots,Y_n)$ and the fact that $-Y_i\overset{d}{=}Y_i$.


Since $U_i\stackrel{iid}{\sim}\cN(0,1)$, $Y_i\stackrel{iid}{\sim}\cN(0,1)$, and $\gamma_0\leq 1$,
we have
\[
\EE[I_2] = \EE\left[\frac{1}{n}\sum_{i=1}^n U_i^2(\gamma_0U_i+2Y_i)(\gamma_0 U_i+Y_i)\right]\leq C
\]
and 
\begin{align*}
\var[I_2]&=\operatorname{Var}\left(\frac{1}{n}\sum_{i=1}^n U_i^2(\gamma_0U_i+2Y_i)(\gamma_0U_i+Y_i)\right)\\
&  =\frac{1}{n^2}\sum_{i=1}^n \operatorname{Var}\left(U_i^2(\gamma_0U_i+2Y_i)(\gamma_0U_i+Y_i)\right) \\
&\qquad+ \frac{1}{n^2}\sum_{i\neq j}\operatorname{Cov}\left(U_i^2(\gamma_0U_i+2Y_i)(\gamma_0U_i+Y_i),U_j^2(\gamma_0U_j+2Y_j)(\gamma_0U_j+Y_j)\right)\\
& \stackrel{(i)}{=}\frac{1}{n^2}\sum_{i=1}^n \operatorname{Var}\left(U_i^2(\gamma_0U_i+2Y_i)(\gamma_0U_i+Y_i)\right) +0\\
&  \leq \frac{C}{n},
\end{align*}
where $(i)$ follows from the independence between $U_i^2(\gamma_0U_i+2Y_i)(\gamma_0U_i+Y_i)$ and $U_j^2(\gamma_0U_j+2Y_j)(\gamma_0U_j+Y_j)$ for $i\neq j$.

By Chebyshev's inequality, we have
\begin{equation}\label{eqn: B-I2}
\PP\left(|I_2-\EE[I_2]|\geq t\right) \leq \frac{\var[I_2]}{t^2}\leq \frac{C}{nt^2}.
\end{equation}
\wp at least $1-C/n$.

Combining \eqref{eqn: B-I_1} and \eqref{eqn: B-I2}, we have \wp 
 at least $1-Ce^{-c\sqrt{d}}-Ce^{-c\log n}$ that
\[
I_1+I_2\leq -c \frac{d\log n}{n} + C.
\]
\qed

\section{Proof of Lemma \ref{thm: strong-convex}}
\label{sec: proof-hessian}

In this proof, we aim to provide a lower bound for $\lambda_{\operatorname{min}}(\nabla^2 L(w^*))$. To this end, we only need to lower bound $u^{\top}\nabla^2 L(w^*) u$ for uniformly all $\|u\|=1$.

For every $\|u\|=1$ and every $N\geq 1$, we have
\begin{equation*}
\begin{aligned}
     u^{\top}\nabla^2 L(w^*) u =\frac{2}{n}\sum_{i=1}^n (u^{\top}x_i)^2(w^{*\top}x_i)^2
    \geq \frac{2}{n}\sum_{i=1}^n (u^{\top}x_i)^2(w^{*\top}x_i)^21_{|w^{*\top}x_i|\leq N}.
\end{aligned}
\end{equation*}
By Lemma \ref{lem: III concentrate}, with probability at least $1-Ce^{Cd-cn/N^2}-2(1+2/\ep)^d e^{-c\operatorname{min}(n\ep^2,n\ep/N^2)}-\frac{C}{\ep^2 n}$ it holds that 
\[
    \sup_{u\in\SS^{d-1}}\left|\frac{1}{n}\sum_{i=1}^{n}(u^{\top}x_i)^2(w^{*\top}x_i)^21_{|w^{*\top}x_i|\leq N}-\EE\left[ (u^{\top}x)^2(w^{*\top}x)^21_{|w^{*\top}x|\leq N}\right]\right|\leq C\ep.
    \]
Moreover, we have
    $
    \EE \left[(u^{\top}x)^2(w^{*\top}x)^21_{|w^{*\top}x|\leq N}\right]\geq 1+2\langle w^*,u\rangle^2 - CN^3e^{-\frac{N^2}{2}}
    $.
Therefore, by choosing $N$ to be some large absolute constant and $\ep$ to be some small absolute constant, \wp at least $1-Ce^{Cd-cn}-\frac{C}{n}$ it holds that
\[
u^{\top}\nabla^2 L(w^*) u\geq -C\ep+2-CN^3e^{-\frac{N^2}{2}} \geq c>0
\]
uniformly for all $\|u\|=1$. 

Plugging $n\geq Cd$ in, we complete the proof.
\qed

\section{Proof of Theorem \ref{thm: one point convex positive}}\label{app:three}
We firstly restate our theorem.
\mainthmthree*

Still let $\delta=w-w^*$. Then, we have
\begin{equation}\label{equa: main decomp in thm 3}
\begin{aligned}
    \inn{\nabla L(w),w-w^*} 
    &=\frac{1}{n}\sum_{i=1}^{n}(\delta^{\top}x_i)^2(\delta^{\top}x_i +2w^{*\top}x_i)(\delta^{\top}x_i +w^{*\top}x_i)\\
    &= \frac{1}{n}\sum_{i=1}^{n}(\delta^{\top}x_i)^4 + \frac{3}{n}\sum_{i=1}^{n}(\delta^{\top}x_i)^3(w^{*\top}x_i)+ \frac{2}{n}\sum_{i=1}^{n}(\delta^{\top}x_i)^2(w^{*\top}x_i)^2\\
    &=  \frac{1}{2n}\sum_{i=1}^{n}(\delta^{\top}x_i)^2\left((\delta^{\top}x_i)^2+6(\delta^{\top}x_i)(w^{*\top}x_i)1_{|w^{*\top}x_i|\leq t}+4(w^{*\top}x_i)^21_{|w^{*\top}x_i|\leq t}\right)\\
    &\quad +\frac{1}{2n}\sum_{i=1}^{n}(\delta^{\top}x_i)^2\left((\delta^{\top}x_i)^2+6(\delta^{\top}x_i)(w^{*\top}x_i)+4(w^{*\top}x_i)^2\right)1_{|w^{*\top}x_i|\geq t} \\
    &\quad + \frac{1}{2n}\sum_{i=1}^{n}(\delta^{\top}x_i)^41_{|w^{*\top}x_i|\geq t}\\
    &\geq \frac{1}{2n}\sum_{i=1}^{n}(\delta^{\top}x_i)^2\left((\delta^{\top}x_i)^2+6(\delta^{\top}x_i)(w^{*\top}x_i)1_{|w^{*\top}x_i|\leq t}+4(w^{*\top}x_i)^21_{|w^{*\top}x_i|\leq t}\right)\\
    &\quad +\frac{1}{2n}\sum_{i=1}^{n}(\delta^{\top}x_i)^2\left((\delta^{\top}x_i)^2+6(\delta^{\top}x_i)(w^{*\top}x_i)+4(w^{*\top}x_i)^2\right)1_{|w^{*\top}x_i|\geq t} \\ 
    &:=A_1 + A_2,
\end{aligned}
\end{equation}
where we drop the last term in the last line and define
\begin{align*}
A_1 &= \frac{1}{2n}\sum_{i=1}^{n}(\delta^{\top}x_i)^2\left((\delta^{\top}x_i)^2+6(\delta^{\top}x_i)(w^{*\top}x_i)1_{|w^{*\top}x_i|\leq t}+4(w^{*\top}x_i)^21_{|w^{*\top}x_i|\leq t}\right)\\
A_2 &= \frac{1}{2n}\sum_{i=1}^{n}(\delta^{\top}x_i)^2\left((\delta^{\top}x_i)^2+6(\delta^{\top}x_i)(w^{*\top} x_i)+4(w^{*\top} x_i)^2\right)1_{|w^{*\top} x_i|\geq t}.
\end{align*}

\paragraph*{\underline{Bound $A_2$}.}
Note that 
    $x^2(x^2+6xy+4y^2)\geq -64 y^4$ for $x\in\RR$ and $y\in \RR$ (Lemma \ref{lemma: preliminary}), we have \wp at least $1-Cte^{\frac{t^2}{2}}/n$,
\begin{align}\label{eqn: C-A2}
   A_2&\geq -\frac{32}{n}\sum_{i=1}^{n}(w^{*\top} x_i)^41_{|w^{*\top} x_i|\geq t}\geq - 64\EE\left[ (w^{*\top}x)^41_{|w^{*\top}x|\geq t}\right] \geq -Ct^3 e^{-\frac{t^2}{2}},
\end{align}
where the second and third steps follow from 
 Lemma \ref{lem: density concentration} and Lemma \ref{lem: moment cutoff}, respectively. 

\paragraph*{\underline{Bound $A_1$}.}
Let $\Bar{\delta}=\frac{\delta}{\norm{\delta}{}}$ and $A_1=I_1+I_2+I_3$ with 
\begin{align*}
I_1 &:= \frac{1}{n}\sum_{i=1}^{n}(\delta^{\top}x_i)^4 =\frac{\norm{\delta}{}^4}{n}\sum_{i=1}^{n}(\Bar{\delta}^{\top}x_i)^4\\ 
    I_2 &:= \frac{1}{n}\sum_{i=1}^{n}(\delta^{\top}x_i)^3(w^{*\top}x_i)1_{|w^{*\top}x_i|\leq t}\\
    &= \underbrace{\frac{\norm{\delta}{}^3}{n}\sum_{i=1}^{n}(\Bar{\delta}^{\top}x_i)^31_{|\Bar{\delta}^{\top}x_i|\leq N}(w^{*\top}x_i)1_{|w^{*\top}x_i|\leq t}}_{I_{2,\leq}}+ \underbrace{\frac{\norm{\delta}{}^3}{n}\sum_{i=1}^{n}(\Bar{\delta}^{\top}x_i)^31_{|\Bar{\delta}^{\top}x_i|\geq N}(w^{*\top}x_i)1_{|w^{*\top}x_i|\leq t}}_{I_{2,\geq}}\\ 
    I_3 &:= \frac{1}{n}\sum_{i=1}^{n}(\delta^{\top}x_i)^2(w^{*\top}x_i)^21_{|w^{*\top}x_i|\leq t} = \frac{\norm{\delta}{}^2}{n}\sum_{i=1}^{n}(\Bar{\delta}^{\top}x_i)^2(w^{*\top}x_i)^21_{|w^{*\top}x_i|\leq t}\\ 
    &= \underbrace{\frac{\|\delta\|^2}{n}\sum_{i=1}^{n}(\bar{\delta}^{\top}x_i)^21_{|\Bar{\delta}^{\top}x_i|\leq N}(w^{*\top}x_i)^21_{|w^{*\top}x_i|\leq t}}_{I_{3,\leq}} 
        + \underbrace{\frac{\|\delta\|^2}{n}\sum_{i=1}^{n}(\bar{\delta}^{\top}x_i)^21_{|\Bar{\delta}^{\top}x_i|\geq N}(w^{*\top}x_i)^21_{|w^{*\top}x_i|\leq t}}_{I_{3,\geq}},
\end{align*}
where $N$ is a large constant to be determined later.


With above notations and estimates, we have \wp at least $1-Cte^{\frac{t^2}{2}}/n$ it holds that 
\begin{align}\label{eqn: C4}
\langle \nabla L(w),w-w^*\rangle &\geq \frac{1}{2}\left(I_1+6I_2+4I_3\right)-Ct^3e^{-\frac{t^2}{2}}.
\end{align}
Noting that 
\begin{align*}
   I_{2,\geq} &=\left|\frac{1}{n}\sum_{i=1}^{n}(\delta^{\top}x_i)^31_{|\Bar{\delta}^{\top}x_i|\geq N}(w^{*\top}x_i)1_{|w^{*\top}x_i|\leq t}\right|\\
    &\geq -\sqrt{\frac{1}{n}\sum_{i=1}^{n}(\delta^{\top}x_i)^4}\sqrt{\frac{1}{n}\sum_{i=1}^{n}(\delta^{\top}x_i)^21_{|\Bar{\delta}^{\top}x_i|\geq N}(w^{*\top}x_i)^21_{|w^{*\top}x_i|\leq t}}\\ 
    &=-\sqrt{I_1}\sqrt{I_{3,\geq}},
\end{align*}
we have
\begin{align}\label{eqn: C3}
\notag I_1+6I_2+4I_3 &= I_1+6I_{2,\geq}+6I_{2,\leq}+4I_3 \\ 
\notag &\geq I_1-6\sqrt{I_1}\sqrt{I_{3,\geq}}+6I_{2,\leq}+4I_3\\ 
\notag &= (\sqrt{I_1}-3\sqrt{I_{3,\geq}})^2-9 I_{3,\geq}+4 I_{3} + 6I_{2,\leq}\\ 
&\geq -9 I_{3,\geq}+4 I_{3} + 6I_{2,\leq}.
\end{align}
The following lemmas provide bounds for each terms in the right hand side of the above inequality.
\begin{lemma}[Bound $I_{2,\leq}$]\label{lem: II_leq concentrate}
For any $N\geq 2$, with probability at least $1-Ce^{Cd-cn}-C/n-C(1+CN^3)^d \operatorname{exp}\left(-c\frac{n}{N^8t^2}\right)$, it holds that 
\[
    \sup_{u\in\SS^{d-1}}\left|\frac{1}{n}\sum_{i=1}^{n}(u^{\top}x_i)^31_{|\Bar{\delta}^{\top}x_i|\leq N}(w^{*\top}x_i)1_{|w^{*\top}x_i|\leq t}\right| \leq C .\quad
\]
\end{lemma}
\begin{lemma}[Bound $I_{3}$]\label{lem: III concentrate}
    For any $\ep>0$, with probability at least $1-Ce^{Cd-cn/t^2}-2(1+2/\ep)^d e^{-c\operatorname{min}(n\ep^2,n\ep/t^2)}-\frac{C}{\ep^2 n}$, it holds  that
    \[
    \sup_{u\in\SS^{d-1}}\left|\frac{1}{n}\sum_{i=1}^{n}(u^{\top}x_i)^2(w^{*\top}x_i)^21_{|w^{*\top}x_i|\leq t}-\EE\left[ (u^{\top}x)^2(w^{*\top}x)^21_{|w^{*\top}x|\leq t}\right]\right|\leq C\ep .
    \]
    Moreover, 
    $
    \EE \left[(u^{\top}x)^2(w^{*\top}x)^21_{|w^{*\top}x|\leq t}\right]\geq 1+2\langle w^*,u\rangle^2 - Ct^3e^{-\frac{t^2}{2}}
    $.
\end{lemma}

\begin{lemma}[Bound $I_{3,\geq}$]\label{lem: III_geq concentrate}
     For any $2\leq N\leq t$ and $\ep>0$, with probability at least $1-Ce^{Cd-cn/t^2}-C\frac{1}{\ep^2 n}-C\left(1+CN^2/\ep\right)^de^{-c\operatorname{min}(n\ep^2,n\ep/t^2)}$ it holds that
    \[
    \sup_{u\in\SS^{d-1}}\frac{1}{n}\sum_{i=1}^{n}(u^{\top}x_i)^21_{|u^{\top}x_i|\geq N}(w^{*\top}x_i)^2 1_{|w^{*\top}x_i|\leq t}\leq C\ep+CN^3e^{-\frac{N^2}{16}} .
    \quad
    \]
\end{lemma}
The proofs of the above three lemmas are deferred to Appendix \ref{sec: proof-I2-leq-con}, \ref{sec: proof-I3-con}, and \ref{sec: proof-I3-geq-con}, respectively.

\paragraph*{Combining all estimates.}
By Lemma \ref{lem: II_leq concentrate},  \ref{lem: III concentrate}, and \ref{lem: III_geq concentrate}, if $N\geq 2$, \wp 
\[
  1- C e^{Cd - cn/t^2} - C(1+CN^3)^d \operatorname{exp}\left(-c\frac{n}{N^8t^2}\right) - C\left(1+\frac{CN^2}{\ep}\right)^d e^{-c\min(c\ep^2,n\ep/t^2)} - C\frac{1}{\ep^2n},
\]
it holds that
\begin{align*}
4I_3-9I_{3,\geq}+6I_{2,\leq}&\geq \norm{\delta}{}^2\left(-C\ep+1+2\langle u,w^*\rangle^2-Ct^3e^{\frac{t^2}{2}}\right)-9\norm{\delta}{}^2\left(C\ep+CN^3e^{-\frac{N^2}{16}}\right)-6C\norm{\delta}{}^3 .
\end{align*}

By taking $t$ and $\ep$ to be smaller than some absolute constants and $N$ to be a large enough absolute constant, we can conclude that \wp 
\begin{equation}\label{eqn: prob}
1-\frac{C}{n} - Ce^{C d - cn/t^2} 
\end{equation}
it holds that
\[
  4I_3-9I_{3,\geq}+6I_{2,\leq}\geq c\|\delta\|^2 - C\|\delta\|^3. 
\]

Combining with \eqref{eqn: C4} and \eqref{eqn: C3}, we have when $\norm{\delta}{}$ is smaller than some absolute constant and $\|\delta\|^2\gtrsim t^3e^{-\frac{t^2}{2}}$, it holds that 
\[
\langle \nabla L(w), w-w^*\rangle \geq c\norm{\delta}{}^2 -Ct^3e^{-\frac{t^2}{2}}\geq c\norm{\delta}{}^2 
\]
\wp at least $1-Cte^{\frac{t^2}{2}}/n-Ce^{-c(n/t^2-Cd)}$. Thus, we complete the proof.
\qed



\subsection{Proof of Lemma \ref{lem: II_leq concentrate}}
\label{sec: proof-I2-leq-con}

\begin{proof}
Let $h(z)=z^3 1_{|z|\leq N}$. Our task is to provide a uniform bound of 
\[
    \frac{1}{n}\sum_{i=1}^{n}(u^{\top}x_i)^31_{|u^{\top}x_i|\leq N}(w^{*\top}x_i) 1_{|w^{*\top}x_i|\leq t} = \frac{1}{n}\sum_{i=1}^{n}h(u^{\top}x_i) 1_{|w^{*\top}x_i|\leq t}.
\]
for any $u\in\SS^{d-1}$. Unfortunately, $h(\cdot)$ is not Lipschitz continuous (see Figure \ref{fig:cube}) and consequently, we cannot apply standard uniform concentration inequalities. To resolve this issue, we define following auxiliary functions:
$$\psi(z)=\left\{\begin{matrix}
  z^3 & |z|\leq N\\
  \text{sgn}(z)N^3(N+1-|z|) & N<|z|\leq N+1\\
  0 & |z|>N+1
\end{matrix}\right.
$$ 
and 
$$\phi(z)=\left\{\begin{matrix}
  0 & |z|\leq N-1\\
  N^3(1-\left||z|-N\right|) & N-1<|z|\leq N+1\\
  0 & |z|>N+1
\end{matrix}\right.,
$$
which satisfies 
$
    \mathrm{Lip}(\psi)=\mathrm{Lip}(\phi)\leq N^3.
$
In Figure  \ref{fig:cube}, we provide a visualization of $\psi$ and $\phi$. 

\begin{figure}[!h]
    \centering
    \includegraphics[width=0.5\textwidth]{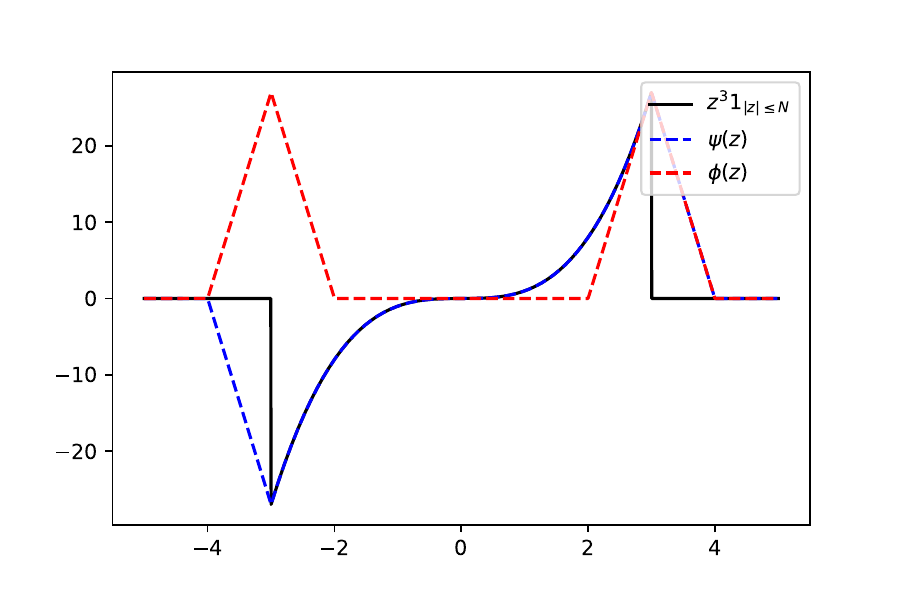}
    \caption{Illustration of $\psi(z)$ and $\phi(z)$ when $N=3$.}
    \label{fig:cube}
\end{figure}

It is easy to verify that $\abs{z^31_{|z|\leq N}-\psi(z)}\leq \phi(z)$, by which we have
\begin{equation}
    \begin{aligned}
    &\left|\frac{1}{n}\sum_{i=1}^{n}(u^{\top}x_i)^31_{|u^{\top}x_i|\leq N}(w^{*\top}x_i) 1_{|w^{*\top}x_i|\leq t}-
    \frac{1}{n}\sum_{i=1}^{n}\psi(u^{\top}x_i)(w^{*\top}x_i) 1_{|w^{*\top}x_i|\leq t}\right| \\
&\quad\quad=\left|\frac{1}{n}\sum_{i=1}^{n}\left((u^{\top}x_i)^31_{|u^{\top}x_i|\leq N}-\psi(u^{\top}x_i)\right)(w^{*\top}x_i) 1_{|w^{*\top}x_i|\leq t}\right|\\
    &\quad\quad\leq \frac{1}{n}\sum_{i=1}^{n}\left|(u^{\top}x_i)^31_{|u^{\top}x_i|\leq N}-\psi(u^{\top}x_i) \right|\abs{w^{*\top}x_i} 1_{|w^{*\top}x_i|\leq t}\\
    &\quad\quad\leq \frac{1}{n}\sum_{i=1}^{n}\phi(u^{\top}x_i)\abs{w^{*\top}x_i} 1_{|w^{*\top}x_i|\leq t}.
\end{aligned}
\end{equation}
Thus, 
\begin{align*}
    \left|\frac{1}{n}\sum_{i=1}^{n}(u^{\top}x_i)^31_{|u^{\top}x_i|\leq N}(w^{*\top}x_i) 1_{|w^{*\top}x_i|\leq t}\right|
    &\leq \left|\frac{1}{n}\sum_{i=1}^{n}\psi(u^{\top}x_i)(w^{*\top}x_i) 1_{|w^{*\top}x_i|\leq t}\right|\\ 
    &\qquad +
    \frac{1}{n}\sum_{i=1}^{n}\phi(u^{\top}x_i)\abs{w^{*\top}x_i} 1_{|w^{*\top}x_i|\leq t},
\end{align*}
where $\psi$ and $\phi$ are both Lipschitz continuous.

By Lemma \ref{lem: lip concentrate}, \wp at least $1-Ce^{Cd-cn}-C/n-C(1+CN^3/\ep)^d \operatorname{exp}\left(-c\frac{n\ep^2}{N^8t^2}\right)$, we have
\[
\sup_{u\in\SS^{d-1}}\left|\frac{1}{n}\sum_{i=1}^{n}\psi(u^{\top}x_i)(w^{*\top}x_i) 1_{|w^{*\top}x_i|\leq t}-\EE \left[\psi(u^{\top}x)(w^{*\top}x) 1_{|w^{*\top}x|\leq t}\right]\right|\leq C\ep,
\]
and
\[
   \sup_{u\in\SS^{d-1}}\left|\frac{1}{n}\sum_{i=1}^{n}\phi(u^{\top}x_i)\abs{w^{*\top}x_i} 1_{|w^{*\top}x_i|\leq t}-\EE\left[ \phi(u^{\top}x)\abs{w^{*\top}x} 1_{|w^{*\top}x|\leq t}\right]\right|\leq C\ep.
\]
Thus, we have \wp $1-Ce^{Cd-cn}-C/n-C(1+CN^3/\ep)^d \operatorname{exp}\left(-c\frac{n\ep^2}{N^8t^2}\right)$, it holds that 
\begin{align}\label{eqn: D0}
\notag \sup_{u\in \SS^{d-1}}&\left|\frac{1}{n}\sum_{i=1}^{n}(u^{\top}x_i)^31_{|u^{\top}x_i|\leq N}(w^{*\top}x_i) 1_{|w^{*\top}x_i|\leq t}\right| \\ 
&\leq\left|\EE \left[\psi(u^{\top}x)(w^{*\top}x) 1_{|w^{*\top}x|\leq t}\right]\right|+ \left|\EE\left[ \phi(u^{\top}x)\abs{w^{*\top}x} 1_{|w^{*\top}x|\leq t}\right]\right|.
\end{align}

Next we control the expectations. First, by the fact  $0\leq\phi(z)\leq  N^31_{|z|\geq N-1}$, we have 
\begin{align}\label{eqn: D1}
\notag    \EE \left[\phi(u^{\top}x)\abs{w^{*\top}x} 1_{|w^{*\top}x|\leq t}\right]
 \notag     &\leq N^3 \EE  \left[1_{\abs{u^{\top}x}\geq N-1}  \left|(w^{*\top}x) 1_{|w^{*\top}x|\leq t}\right|\right]\\
\notag      &\leq N^3 \sqrt{\EE\left[ 1_{\abs{u^{\top}x}\geq N-1}\right]}\sqrt{\EE\left[ (w^{*\top}x)^2 1_{|w^{*\top}x|\leq t}\right]}\\
    &\lesssim N^3 e^{-\frac{(N-1)^2}{4}},
\end{align}
where the second and third step follow from the Cauchy–Schwarz inequality and Lemma \ref{lem: moment cutoff}, respectively. 
Furthermore, for any $u\in\SS^{d-1}$, we have 
\begin{align}\label{eqn: D2}
\notag    \left|\EE\left[ \psi(u^{\top}x)(w^{*\top}x) 1_{|w^{*\top}x|\leq t}\right]\right|
&\leq \sqrt{\EE \left[\psi^2(u^{\top}x)\right]}\sqrt{\EE\left[ (w^{*\top}x)^2 1_{|w^{*\top}x|\leq t}\right]}\\
\notag &\leq  \EE_{z\sim \cN(0,1)} \left[z^6\right] \EE_{z\sim \cN(0,1)} \left[z^2\right]\\
&\leq C .
\end{align}
By plugging \eqref{eqn: D1} and \eqref{eqn: D2} into \eqref{eqn: D0}  and taking $\ep=1$, we have \wp at least $1-Ce^{Cd-cn}-C/n-C(1+CN^3)^d \operatorname{exp}\left(-c\frac{n}{N^8t^2}\right)$ it holds that 
\[
     \sup_{u\in \SS^{d-1}}\left|\frac{1}{n}\sum_{i=1}^{n}(u^{\top}x_i)^31_{|u^{\top}x_i|\leq N}(w^{*\top}x_i) 1_{|w^{*\top}x_i|\leq t}\right|\leq C.
\]

\end{proof}

\subsection{Proof of Lemma \ref{lem: III concentrate}}
\label{sec: proof-I3-con}

    Consider the decomposition $u=\inn{u,w^*}w^*+u^\perp$ where $\inn{u^\perp,w^*}=0$. Then 
    \begin{align}\label{eqn: D5}
  \notag      &\frac{1}{n}\sum_{i=1}^{n}(u^{\top}x_i)^2(w^{*\top}x_i)^21_{|w^{*\top}x_i|\leq t}\\
  \notag        &\quad\quad=\inn{u,w^*}^2 \frac{1}{n}\sum_{i=1}^{n}(w^{*\top}x_i)^41_{|w^{*\top}x_i|\leq t} + 2\inn{u,w^*}\frac{1}{n}\sum_{i=1}^{n}((u^{\perp})^{\top}x_i)(w^{*\top}x_i)^31_{|w^{*\top}x_i|\leq t}\\
        &\quad\quad\quad +\frac{1}{n}\sum_{i=1}^{n}((u^{\perp})^{\top}x_i)^2(w^{*\top}x_i)^21_{|w^{*\top}x_i|\leq t} .
    \end{align}

We bound the first term by  Markov's inequality. \wp at least $1-\frac{C}{n\ep^2}$, we have
\begin{align}\label{eqn: D6}
\notag &\sup_{u\in\SS^{d-1}}\abs{\inn{u,w^*}^2 \frac{1}{n}\sum_{i=1}^{n}(w^{*\top}x_i)^41_{|w^{*\top}x_i|\leq t}-\EE\left[\inn{u,w^*}^2 (w^{*\top}x)^41_{|w^{*\top}x|\leq t}\right]} \\ 
\notag   &\quad\quad\leq \sup_{u\in\SS^{d-1}} \inn{u,w^*}^2\abs{\frac{1}{n}\sum_{i=1}^{n}(w^{*\top}x_i)^41_{|w^{*\top}x_i|\leq t}-\EE\left[ (w^{*\top}x)^41_{|w^{*\top}x|\leq t}\right]} \\
&\quad\quad\leq \abs{\frac{1}{n}\sum_{i=1}^{n}(w^{*\top}x_i)^41_{|w^{*\top}x_i|\leq t}-\EE\left[ (w^{*\top}x)^41_{|w^{*\top}x|\leq t}\right]} \leq C\ep .
\end{align}

We next bound the second term. By using Lemma \ref{lem: III concentrate part 2}, we have  \wp at least $(1-Ce^{Cd-cn}-2(1+2/\ep)^d\operatorname{exp}(-c\ep^2 n))(1-C\frac{1}{\ep^2 n})$ it holds that
\begin{equation}\label{eqn: D8}
\sup_{u\in\SS^{d-1}}\abs{\frac{1}{n}\sum_{i=1}^{n}((u^{\perp})^{\top}x_i)(w^{*\top}x_i)^31_{|w^{*\top}x_i|\leq t}-\EE\left[((u^{\perp})^{\top}x)(w^{*\top}x)^31_{|w^{*\top}x|\leq t}\right]} \leq C\ep.
\end{equation}

As for the third term, it always holds that $0\leq (w^{*\top}x_i)^2 1_{\abs{w^{*\top}x_i}\leq t} \leq t^2 $. Hence, by Lemma \ref{lem: III concentrate part 3}, we have
\begin{align}\label{eqn: D7}
\sup_{u\in\SS^{d-1}}\abs{\frac{1}{n}\sum_{i=1}^{n}(w^{*\top}x_i)^21_{|w^{*\top}x_i|\leq t}((u^{\perp})^{\top}x_i)^2 - \EE\left[(w^{*\top}x)^21_{|w^{*\top}x|\leq t}((u^{\perp})^{\top}x)^2\right]} \leq C\ep
\end{align}
\wp at least $(1-Ce^{Cd-cn/t^2}-2\left(1+2/\ep\right)^d\operatorname{exp}\left(-c\operatorname{min}(n\ep^2,n\ep/t^2)\right))(1-C\frac{1}{\ep^2n})$.

Plugging \eqref{eqn: D6}, \eqref{eqn: D8}, and \eqref{eqn: D7} into \eqref{eqn: D5}, we complete the proof of the uniform concentration part.

Lastly, we turn to  lower bound the expectation $\EE\left[ (u^{\top}x)^2(w^{*\top}x)^21_{|w^{*\top}x|\leq t}\right]$ by
 \begin{align*}
        &\EE \left[(u^{\top}x)^2(w^{*\top}x)^21_{|w^{*\top}x|\leq t}\right]\\
        &\quad\quad=\langle w^*,u\rangle^2 \EE\left[(w^{*\top}x)^41_{|w^{*\top}x|\leq t}\right] + 2\langle w^*,u\rangle\EE\left[(w^{*\top}x)^31_{|w^{*\top}x|\leq t}((u^{\perp})^{\top}x)\right]\\
        &\quad\quad\quad +\EE\left[(w^{*\top}x)^21_{|w^{*\top}x|\leq t}((u^{\perp})^{\top}x)^2\right]\\
        &\quad\quad=\langle w^*,u\rangle^2 \EE\left[(w^{*\top}x)^41_{|w^{*\top}x|\leq t}\right] +(1-\langle w^*,u\rangle^2)\EE\left[(w^{*\top}x)^21_{|w^{*\top}x|\leq t}\right]\\
        &\quad\quad= 1+2\langle w^*,u\rangle^2 -\langle w^*,u\rangle^2 \EE\left[(w^{*\top}x)^41_{|w^{*\top}x|\geq t}\right] - (1-\langle w^*,u\rangle^2)\EE\left[(w^{*\top}x)^21_{|w^{*\top}x|\geq t}\right]\\
        &\quad\quad\geq 1+2\langle w^*,u\rangle^2 - \EE\left[(w^{*\top}x)^41_{|w^{*\top}x|\geq t}\right] - \EE\left[(w^{*\top}x)^21_{|w^{*\top}x|\geq t}\right]\\
        &\quad\quad\geq 1+2\langle w^*,u\rangle^2 - Ct^3e^{-\frac{t^2}{2}},
    \end{align*}
where the second step use the independence between $(w^{*})^{\top}x$ and $(u^{\perp})^{\top}x$ for $x\sim\cN(0,1)$ and $\EE[(u^{\perp})^{\top}x]=0$; the third step uses $\EE_{z\sim\cN(0,1)}[z^4]=3$; the fourth step uses $ 0\leq \langle w^*,u\rangle^2\leq 1$; the last step uses the tail bound of $\cN(0,1)$ given by Lemma \ref{lem: moment cutoff}.

\qed

\subsection{Proof of Lemma \ref{lem: III_geq concentrate}}
\label{sec: proof-I3-geq-con}

\begin{figure}[!h]
    \centering
    \includegraphics[width=0.45\textwidth]{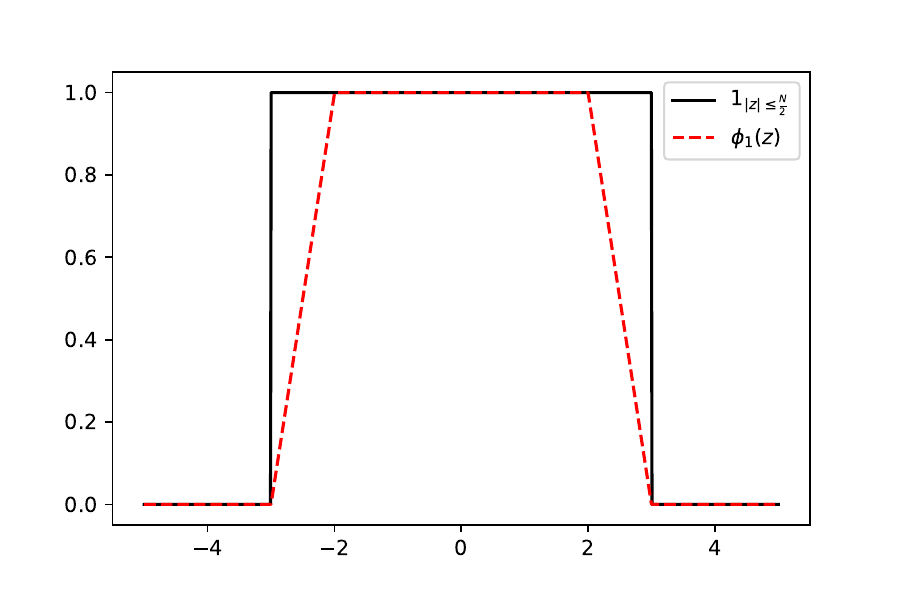}
    \hspace*{-2em}
    \includegraphics[width=0.45\textwidth]{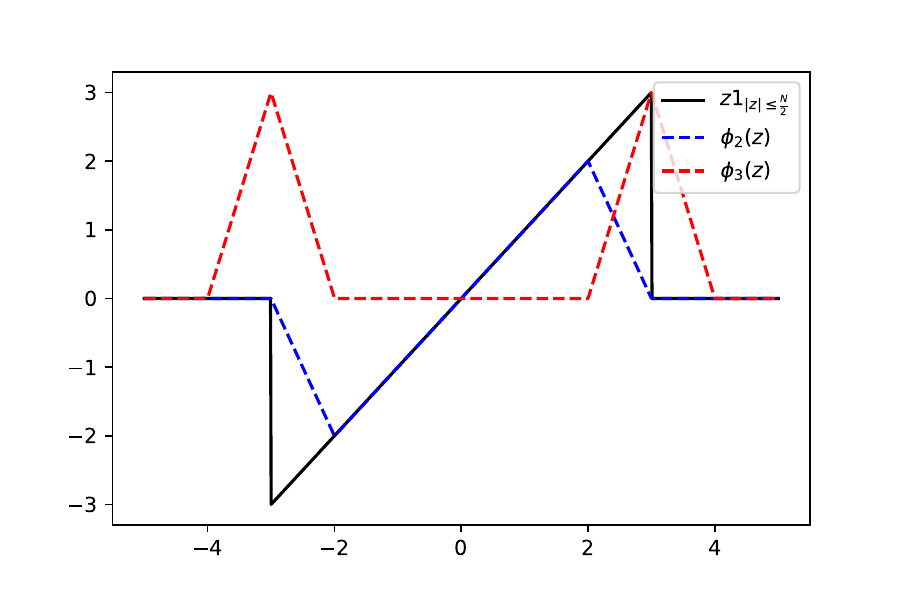}
    \hspace*{-2em}
    \includegraphics[width=0.45\textwidth]{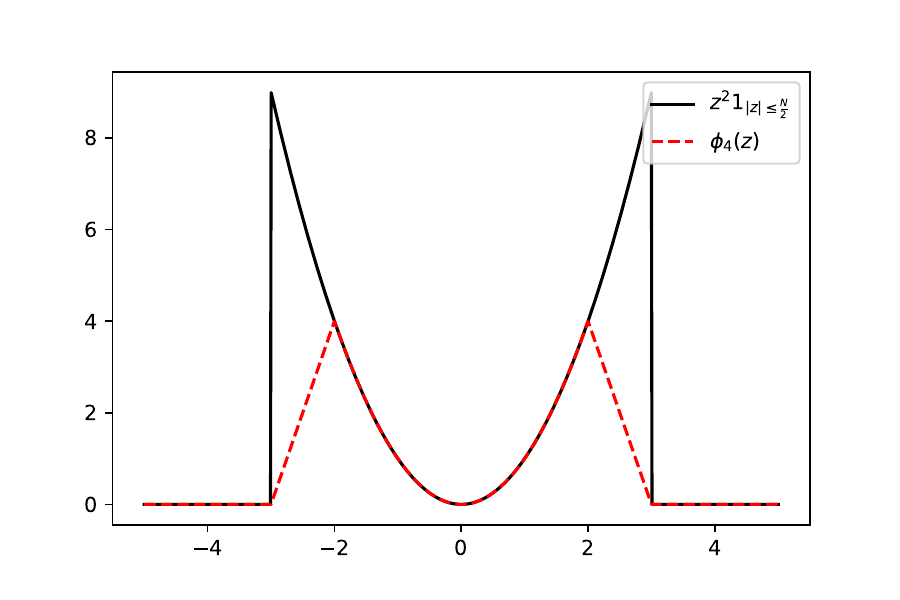}
    \caption{
        An illustration of the auxiliary functions $\phi_1,\phi_2,\phi_3$, and $\phi_4$ for $N=6$. It is obvious that for any $z\in\RR$,  $1_{|z|\leq N/2}\geq \phi_1(z)\geq 1_{|z|\leq N/2-1}$
        and $z^2 1_{|z|\leq N/2}\geq \phi_3(z)$.
    }
    \label{fig:phi}
\end{figure}

We first lower bound the following quantity 
\[
\frac{1}{n}\sum_{i=1}^{n}(u^{\top}x_i)^21_{|u^{\top}x_i|\leq N}(w^{*\top}x_i)^2 1_{|w^{*\top}x_i|\leq t}.
\]
    Write $u=\inn{u,w^*}w^*+u^\perp$, where $\inn{u^\perp,w^*}=0$. 
    Note that 
\begin{equation}\label{eqn: D9}
1_{|u^{\top}x_i|\leq N}\geq 1_{|(u^{\perp})^{\top}x_i|\leq \frac{N}{2}}1_{|\inn{u,w^*}w^{*\top}x_i|\leq \frac{N}{2}}\geq 1_{|(u^{\perp})^{\top}x_i|\leq \frac{N}{2}}1_{|w^{*\top}x_i|\leq \frac{N}{2}},
\end{equation}
where the first step is due to that $|(u^{\perp})^{\top}x_i|\leq \frac{N}{2}$ and $|\inn{u,w^*}w^{*\top}x_i|\leq \frac{N}{2}$ can imply $|u^{\top}x_i|\leq N$ by the triangle inequality; the second step is because $|\inn{u,w^*}w^{*\top}x_i|\leq \frac{N}{2}$ implies $|w^{*\top}x_i|\leq \frac{N}{2}$ as $|\inn{u,w^*}|\leq 1$.
Furthermore, we define some smoothed functions that we will use later. 
    \begin{align*}
   \phi_1(z)&=\left\{\begin{matrix}
    1  & |z|\leq\frac{N}{2}-1 \\
    \frac{N}{2}-|z| & \frac{N}{2}-1\leq |z|\leq\frac{N}{2}\\
    0 & |z|\geq\frac{N}{2}
    \end{matrix}\right. \\ 
    \phi_2(z)&=\left\{\begin{matrix}
z & |z|\leq\frac{N}{2}-1 \\
\text{sgn}(z)(\frac{N}{2}-1)(\frac{N}{2}-|z|)  & \frac{N}{2}-1\leq |z|\leq\frac{N}{2}\\
0 & |x|\geq\frac{N}{2}
\end{matrix}\right.\\ 
 \phi_3(z)&=\left\{\begin{matrix}
    0 & |z|\leq \frac{N}{2}-1\\
    \frac{N}{2}(1-||z|-\frac{N}{2}|) & \frac{N}{2}-1<|z|\leq \frac{N}{2}+1\\
    0 & |z|>\frac{N}{2}+1
    \end{matrix}\right.\\
\phi_4(z)&=\left\{\begin{matrix}
    z^2 & |z|\leq \frac{N}{2}-1\\
    (\frac{N}{2}-1)^2(\frac{N}{2}-|z|) & \frac{N}{2}-1<|z|\leq \frac{N}{2}\\
    0 & |z|>\frac{N}{2}
    \end{matrix}\right. .\\
 \end{align*}
For a better understanding of these auxiliary functions, we refer to the visualization in Figure \ref{fig:phi}. Note that $\mathrm{Lip}(\phi_4)\leq CN^2$ and $\mathrm{Lip}(\phi_i)\leq CN$ for $i=1,2,3$.
With all these preparations, we can now deal with the truncation in $I_{3,\geq}$. 

For $t\geq N/2$, we have
\begin{equation}\label{eqn: D11}
    \begin{aligned}
        &\frac{1}{n}\sum_{i=1}^{n}(u^{\top}x_i)^21_{|u^{\top}x_i|\leq N}(w^{*\top}x_i)^21_{|w^{*\top}x_i|\leq t}\\
        &\geq \frac{1}{n}\sum_{i=1}^{n}(u^{\top}x_i)^21_{|(u^{\perp})^{\top}x_i|\leq \frac{N}{2}}1_{|w^{*\top}x_i|\leq \frac{N}{2}}(w^{*\top}x_i)^21_{|w^{*\top}x_i|\leq t}\\
        &=\frac{1}{n}\sum_{i=1}^{n}(u^{\top}x_i)^21_{|(u^{\perp})^{\top}x_i|\leq \frac{N}{2}}(w^{*\top}x_i)^21_{|w^{*\top}x_i|\leq \frac{N}{2}}\\
        &=\inn{u,w^*}^2 \frac{1}{n}\sum_{i=1}^{n}(w^{*\top}x_i)^41_{|w^{*\top}x_i|\leq \frac{N}{2}}1_{|(u^{\perp})^{\top}x_i|\leq \frac{N}{2}} + 2\inn{u,w^*}\frac{1}{n}\sum_{i=1}^{n}(w^{*\top}x_i)^31_{|w^{*\top}x_i|\leq \frac{N}{2}}((u^{\perp})^{\top}x_i)1_{|(u^{\perp})^{\top}x_i|\leq \frac{N}{2}}\\
        &\quad +\frac{1}{n}\sum_{i=1}^{n}(w^{*\top}x_i)^21_{|w^{*\top}x_i|\leq \frac{N}{2}}((u^{\perp})^{\top}x_i)^21_{|(u^{\perp})^{\top}x_i|\leq \frac{N}{2}}\\
        &\geq \inn{u,w^*}^2 \frac{1}{n}\sum_{i=1}^{n}(w^{*\top}x_i)^41_{|w^{*\top}x_i|\leq \frac{N}{2}}\phi_1((u^{\perp})^{\top}x_i) + 2\inn{u,w^*}\frac{1}{n}\sum_{i=1}^{n}(w^{*\top}x_i)^31_{|w^{*\top}x_i|\leq \frac{N}{2}}\phi_2((u^{\perp})^{\top}x_i)\\
        &\quad+ 2\inn{u,w^*}\frac{1}{n}\sum_{i=1}^{n}(w^{*\top}x_i)^31_{|w^{*\top}x_i|\leq \frac{N}{2}}\left(((u^{\perp})^{\top}x_i)1_{|(u^{\perp})^{\top}x_i|\leq \frac{N}{2}}-\phi_2((u^{\perp})^{\top}x_i)\right)\\
        &\quad +\frac{1}{n}\sum_{i=1}^{n}(w^{*\top}x_i)^21_{|w^{*\top}x_i|\leq \frac{N}{2}}\phi_4((u^{\perp})^{\top}x_i) .\\
    \end{aligned},
\end{equation}
where the first step
    follows from \eqref{eqn: D9}; the last step uses the fact that $1_{\abs{z}\leq \frac{N}{2}}\geq \phi_1(z)$ and  $z^21_{\abs{z}\leq \frac{N}{2}}\geq \phi_4(z)$ for any $z\in\RR$ (we refer to Figure \ref{fig:phi} to see why these hold).
   
Next we shall bound the three terms in \eqref{eqn: D11} separately. 

\paragraph*{Bound the first term.} By $\mathrm{Lip}(\phi_1)\leq CN$ and 
    Lemma \ref{lem: III concentrate part 2}, we have \wp at least 
    \[
    p_1=(1-Ce^{Cd-cn}-2(1+2CN/\ep)^d\operatorname{exp}(-c\ep^2 n))(1-C\frac{1}{\ep^2 n}), 
    \]
    it holds that
    \[
    \sup_{u\in\SS^{d-1}}\left|\frac{1}{n}\sum_{i=1}^{n}(w^{*\top}x_i)^41_{|w^{*\top}x_i|\leq \frac{N}{2}}\phi_1((u^{\perp})^{\top}x_i)-\EE\left[(w^{*\top}x)^41_{|w^{*\top}x|\leq \frac{N}{2}}\phi_1((u^{\perp})^{\top}x)\right]\right|\leq C\ep
    \]
    where 
\begin{align*}
    \EE \left[(w^{*\top}x)^41_{|w^{*\top}x|\leq \frac{N}{2}}\phi_1((u^{\perp})^{\top}x)\right]
    &= \EE\left[ (w^{*\top}x)^41_{|w^{*\top}x|\leq \frac{N}{2}}\right] \EE\left[ \phi_1((u^{\perp})^{\top}x)\right]\\
    &\geq \EE \left[(w^{*\top}x)^41_{|w^{*\top}x|\leq \frac{N}{2}}\right] \EE \left[1_{|(u^{\perp})^{\top}x|\leq \frac{N}{2}-1}\right]\\
    &= \left(3-\EE\left[(w^{*\top}x)^41_{|w^{*\top}x|\geq \frac{N}{2}}\right]\right)\left(1-\EE \left[1_{|(u^{\perp})^{\top}x|\geq \frac{N}{2}-1}\right]\right)\\
    &\geq (3-CN^3e^{-\frac{N^2}{8}})(1-Ce^{-\frac{N^2}{16}})
\end{align*}
The first step is due to $\phi_1(z)\geq 1_{|z|\leq \frac{N}{2}-1}$ for any $z\in\RR$ and the last step follows from Lemma \ref{lemma: tail of gaussian} and Lemma \ref{lem: moment cutoff}.

\paragraph*{Bound the second term.} By $\mathrm{Lip}(\phi_2)\leq CN$ and Lemma \ref{lem: III concentrate part 2}, \wp at least 
\[
p_2=(1-Ce^{Cd-cn}-2(1+2CN/\ep)^d\operatorname{exp}(-c\ep^2 n))(1-C\frac{1}{\ep^2 n}), 
\]
we have 
    \[
    \sup_{u\in\SS^{d-1}}\left|\frac{1}{n}\sum_{i=1}^{n}(w^{*\top}x_i)^31_{|w^{*\top}x_i|\leq \frac{N}{2}}\phi_2((u^{\perp})^{\top}x_i)-\EE \left[(w^{*\top}x)^31_{|w^{*\top}x|\leq \frac{N}{2}}\phi_2((u^{\perp})^{\top}x)\right]\right|\leq C\ep,
    \]
where 
\begin{align*}
    \EE \left[(w^{*\top}x)^31_{|w^{*\top}x|\leq \frac{N}{2}}\phi_2((u^{\perp})^{\top}x)\right]=0 .
\end{align*}

\paragraph*{Bound the third term.} 
First, note that
\begin{align*}
    &\left|\frac{1}{n}\sum_{i=1}^{n}(w^{*\top}x_i)^31_{|w^{*\top}x_i|\leq \frac{N}{2}}\left(((u^{\perp})^{\top}x_i)1_{|(u^{\perp})^{\top}x_i|\leq \frac{N}{2}}-\phi_2((u^{\perp})^{\top}x_i)\right)\right|\\ 
    &\qquad\qquad \leq \frac{1}{n}\sum_{i=1}^{n}|w^{*\top}x_i|^31_{|w^{*\top}x_i|\leq \frac{N}{2}}\phi_3((u^{\perp})^{\top}x_i)
\end{align*}
By $\mathrm{Lip}(\phi_3)\leq CN$  and Lemma \ref{lem: III concentrate part 2}, \wp at least 
\[
p_3=(1-Ce^{Cd-cn}-2(1+CN/\ep)^d\operatorname{exp}(-c\ep^2 n))(1-C\frac{1}{\ep^2 n}),
\] 
it holds that 
\[
\sup_{u\in\SS^{d-1}}\abs{\frac{1}{n}\sum_{i=1}^{n}|w^{*\top}x_i|^31_{|w^{*\top}x_i|\leq \frac{N}{2}}\phi_3((u^{\perp})^{\top}x_i)-\EE\left[|w^{*\top}x|^31_{|w^{*\top}x|\leq \frac{N}{2}}\phi_3((u^{\perp})^{\top}x)\right]}\leq C\ep,
\]
where
\begin{align*}
    \EE \left[|w^{*\top}x|^31_{|w^{*\top}x|\leq \frac{N}{2}}\phi_3((u^{\perp})^{\top}x)\right]
    &= \EE\left[ |w^{*\top}x|^31_{|w^{*\top}x|\leq \frac{N}{2}}\right] \EE\left[ \phi_3((u^{\perp})^{\top}x_i)\right]\\
    &\lesssim \EE\left[ \phi_3((u^{\perp})^{\top}x)\right]\\
    &\lesssim e^{-\frac{N^2}{16}} .
\end{align*}  

\paragraph*{Bound the fourth term.} By $\mathrm{Lip}(\phi_4)\leq CN^2$ and Lemma \ref{lem: III concentrate part 2}, we have  \wp at least 
\[
p_4=(1-Ce^{Cd-cn}-2(1+2CN^2/\ep)^d\operatorname{exp}(-c\ep^2 n))(1-C\frac{1}{\ep^2 n})
\]
it holds that 
    \[
    \sup_{u\in\SS^{d-1}}\left|\frac{1}{n}\sum_{i=1}^{n}(w^{*\top}x_i)^21_{|w^{*\top}x_i|\leq \frac{N}{2}}\phi_4((u^{\perp})^{\top}x_i)-\EE \left[(w^{*\top}x)^21_{|w^{*\top}x|\leq \frac{N}{2}}\phi_4((u^{\perp})^{\top}x)\right]\right|\leq C\ep
    \]
    Let $u^\perp=\|u^\perp\| \overline{u^\perp}$. Noting $\|u^\perp\|\leq 1$, the expectation can be lower bounded as follows
\begin{align*}
    \EE \left[(w^{*\top}x)^21_{|w^{*\top}x|\leq \frac{N}{2}}\phi_4((u^{\perp})^{\top}x)\right]
    &= \EE \left[(w^{*\top}x)^21_{|w^{*\top}x|\leq \frac{N}{2}}\right] \EE\left[ \phi_4((u^{\perp})^{\top}x)\right]\\
    &\geq \EE \left[(w^{*\top}x)^21_{|w^{*\top}x|\leq \frac{N}{2}}\right] \EE \left[((u^{\perp})^{\top}x)^21_{|(u^{\perp})^{\top}x|\leq \frac{N}{2}-1}\right]\\
    &= \|u^\perp\|^2 \EE\left[ (w^{*\top}x)^21_{|w^{*\top}x|\leq \frac{N}{2}}\right] \EE \left[\left(\left(\overline{u^\perp}\right)^{\top}x\right)^21_{|(u^{\perp})^{\top}x|\leq \frac{N}{2}-1}\right]\\
     &\geq \|u^\perp\|^2 \EE\left[ (w^{*\top}x)^21_{|w^{*\top}x|\leq \frac{N}{2}}\right] \EE \left[\left(\left(\overline{u^\perp}\right)^{\top}x\right)^21_{|(\overline{u^\perp})^{\top}x|\leq \frac{N}{2}-1}\right]\\
     &= \|u^\perp\|^2 (1-\EE\left[ (w^{*\top}x_i)^21_{|w^{*\top}x_i|\geq \frac{N}{2}}\right])(1-\EE\left[ \left(\left(\overline{u^\perp}\right)^{\top}x\right)^21_{|(\overline{u^\perp})^{\top}x_i|\geq \frac{N}{2}-1}\right])\\
    &\geq \|u^\perp\|^2(1-CNe^{-\frac{N^2}{16}})^2 .
\end{align*}
The first step uses that $\phi_4(z)\geq z^21_{|z|\leq \frac{N}{2}-1}$ for any $z\in\RR$. 
The second step uses that $1_{|(u^{\perp})^{\top}x|\leq \frac{N}{2}-1}\geq 1_{|(\overline{u^\perp})^{\top}x|\leq \frac{N}{2}-1}$ as $\|u^\perp\|\leq 1$. 
The last step follows from the tail bound of standard normal random variables provided in Lemma \ref{lem: moment cutoff}.

\paragraph*{Combining all estimates.}
Combining all the estimates above, we have \wp at least $\sum_{1\leq i \leq 4} p_i$ it holds that
\begin{align*}
        &\frac{1}{n}\sum_{i=1}^{n}(u^{\top}x_i)^21_{|u^{\top}x_i|\leq N}(w^{*\top}x_i)^21_{|w^{*\top}x_i|\leq t}\\
        & \geq -C\ep+\langle u,w^*\rangle^2\left(3-CN^3e^{-\frac{N^2}{8}}\right)\left(1-Ce^{-\frac{N^2}{16}}\right)+\norm{u^{\perp}}{}^2 \left(1-CNe^{-\frac{N^2}{16}}\right)^2-Ce^{-\frac{N^2}{16}}\\
        & \geq -C\ep + 1+ 2\langle u,w^*\rangle ^2 -CN^3e^{-\frac{N^2}{16}} .
\end{align*}

In addition, by Lemma \ref{lem: III concentrate}, it holds 
\[
   \sup_{u\in\SS^{d-1}} \left|\frac{1}{n}\sum_{i=1}^{n}(u^{\top}x_i)^2(w^{*\top}x_i)^21_{|w^{*\top}x_i|\leq t}-\EE\left[ (u^{\top}x)^2(w^{*\top}x)^21_{|w^{*\top}x|\leq t}\right]\right|\leq C\ep
\]
    \wp at least $1-Ce^{Cd-cn/t^2}-2(1+2/\ep)^d e^{-c\operatorname{min}(n\ep^2,n\ep/t^2)}-C\frac{1}{\ep^2 n}$.

Therefore, we have 
\begin{align*}
    &\frac{1}{n}\sum_{i=1}^{n}(u^{\top}x_i)^21_{|u^{\top}x_i|\geq N}(w^{*\top}x_i)^21_{|w^{*\top}x_i|\leq t}\\
    &= \frac{1}{n}\sum_{i=1}^{n}(u^{\top}x_i)^2(w^{*\top}x_i)^21_{|w^{*\top}x_i|\leq t} - \frac{1}{n}\sum_{i=1}^{n}(u^{\top}x_i)^21_{|u^{\top}x_i|\leq N}(w^{*\top}x_i)^21_{|w^{*\top}x_i|\leq t}\\
    &\leq  C\ep+1+2\langle u,w^*\rangle^2 - \frac{1}{n}\sum_{i=1}^{n}(u^{\top}x_i)^21_{|u^{\top}x_i|\leq N}(w^{*\top}x_i)^21_{|w^{*\top}x_i|\leq t}\\
    &\leq C\ep+CN^3e^{-\frac{N^2}{16}} .
\end{align*}

\qed

\section{Auxiliary Lemmas for Appendix \ref{app:three}}\label{appendix: auxiliary}
\begin{lemma}\label{lemma: preliminary}
For every $x,y\in\RR$, we have
    \[
    x^2(x^2+6xy+4y^2)\geq -64 y^4 .
    \]
\end{lemma}
\begin{proof}
    Let $f_y(x)=x^2(x^2+6xy+4y^2)$. For any fixed $y\in\RR$,  since $\lim_{x\to\infty} f_y(x)=+\infty$ and $f_y(x)$ is continuous, $f_y(x)$ attains its minimum.
    Next, we calculate\[
    \frac{\mathrm{d} f_y}{\mathrm{d} x}= 2x(2x^2+9xy+4y^2).
    \]
    Solving $\frac{\mathrm{d} f_y}{\mathrm{d} x}= 0$ gives $x=ky$, where $k=0$, $k=-\frac{1}{2}$, or $k=-4$.  By comparing the value of $f_y(x)$ at $k=0$, $k=-\frac{1}{2}$, and $k=-4$, one can find that the minimum of $f_y(x)$ is attained at $x=-4y$, and the minimum value of $f_y(x)$ equals to $-64y^4$. 
\end{proof}

\begin{lemma}\label{lem: operator norm stronger}
    Suppose $a=(a_1,\dots,a_n)\in\RR^n$ satisfies $\|a\|_2 \leq C\sqrt{n}$ and $\|a\|_\infty \leq Ct $. Let $X_1,\dots,X_n$ be \iid $\cN(0,I_d)$ random variables.
    For any $\ep>0$, with probability at least $1-Ce^{Cd-cn\operatorname{min}(\ep^2,\ep/t)}$, we have
    \[
    \sup_{u,v\in \SS^{d-1}}\left|\frac{1}{n}\sum_{i=1}^n a_i(u^{\top}X_i)(v^{\top}X_i)-\left(\frac{1}{n}\sum_{i=1}^n a_i\right)u^{\top}v \right|\leq \ep .
    \]
\end{lemma}
\begin{proof}
Let 
    $A:=\frac{1}{n}\sum_{i=1}^n a_i(X_iX_i^{\top}-I)$, and a simple calculation yields
    \[
    \frac{1}{n}\sum_{i=1}^n a_i(u^{\top}X_i)(v^{\top}X_i)-\left(\frac{1}{n}\sum_{i=1}^n a_i\right)u^{\top}v = u^{\top}Av.
    \]
   For any $\delta\in (0,1/2)$, let $S_\delta$ be a $\delta$-net  of $\SS^{d-1}$ with respect to the $\ell^2$ distance. Note that we can choose a $S_{\delta}$ satisfying $\text{card}(S_\delta)\leq (1+\frac{2}{\delta})^{d}$ (see, e.g.,~\citet[Corollary 4.2.13]{vershynin2018high}). Then,~\citet[Exercise 4.4.3]{vershynin2018high} gives
    \[
    \sup_{u,v\in \SS^{d-1}} u^{\top}Av\leq \frac{1}{1-2\delta}\sup_{u,v\in S_\delta}u^{\top} Av .
    \] 
    Therefore, we only need to upper bound the RHS of the above equation.
    
    By Bernstein’s inequality (Lemma \ref{lem: Berstein}), for each $u,v\in \SS^{d-1}$ and any $\ep>0$, we have 
    \[\PP \left(\left|\frac{1}{n}\sum_{i=1}^n a_i(u^{\top}X_i)(v^{\top}X_i)-\left(\frac{1}{n}\sum_{i=1}^n a_i\right)u^{\top}v \right|\geq \ep/2\right)\leq 2\exp{\left(-c\operatorname{min}\left(n\ep^2,\frac{n\ep}{t}\right)\right)}.
    \]
    Next, we derive the following
    \[\PP \left(\sup_{u,v\in S_{\delta}}\left|\frac{1}{n}\sum_{i=1}^n a_i(u^{\top}X_i)(v^{\top}X_i)-\left(\frac{1}{n}\sum_{i=1}^n a_i\right)u^{\top}v \right|\geq \ep/2\right)\leq 2\left(1+\frac{2}{\delta}\right)^{d}\exp{\left(-c\operatorname{min}\left(n\ep^2,\frac{n\ep}{t}\right)\right)}\]
    by taking a union bound over $S_\delta$.
    Now, we choose $\delta=\frac{1}{4}$. Plug that in, and we have \wp at least $1-Ce^{Cd-c\operatorname{min}(n\ep^2,n\ep/t)}$ it holds
    \[
    \sup_{u,v\in \SS^{d-1}}\left|\frac{1}{n}\sum_{i=1}^n a_i(u^{\top}X_i)(v^{\top}X_i)-\left(\frac{1}{n}\sum_{i=1}^n a_i\right)u^{\top}v \right|\leq \ep .
    \]
\end{proof}
\begin{lemma}\label{lem: III concentrate part 2}
    Suppose $f:\RR\to\RR$ is a Lipschitz continuous function such that 
    \[|f(x)-f(\Tilde{x})|\leq L |x-\Tilde{x}|,\quad \forall x,\Tilde{x}\in\RR.\]
    and $\sup_{0\leq a\leq 1}\|f(aZ)\|_{\psi_2} \leq C$ for a standard Gaussian $Z\sim \cN(0,1)$. 
    Let $X_1,\dots,X_n$ be \iid $\cN(0,I_d)$ generated random variables, and let
    $Y_1,\dots,Y_n$ be \iid random variables with $\EE\left[Y^4\right] \leq C$ that is independent of the $X_1,\dots,X_n$.
    
    Then for any $\ep>0$, with probability at least $(1-Ce^{Cd-cn}-2(1+2L/\ep)^d\operatorname{exp}(-c\ep^2 n))(1-C\frac{1}{\ep^2 n})$ it holds 
    \[
    \operatorname{sup}_{\norm{u}{}\leq 1}\abs{\frac{1}{n}\sum_{i=1}^n f(u^{\top}X_i)Y_i-\EE_{Z\sim\cN(0,1)}\left[f(\norm{u}{}Z)\right]\EE\left[Y\right]} \leq C\ep .
    \]
    
\end{lemma}
\begin{proof}
For any $\ep>0$, using standard Markov's inequality argument, we have \wp at least $1-C\frac{1}{\ep^2 n}$,
\[
\abs{\frac{1}{n}\sum_{i=1}^n Y_i -\EE\left[Y\right]} \leq \ep \text{  and  } \abs{\frac{1}{n}\sum_{i=1}^n Y_i^2 -\EE\left[Y^2\right]} \leq \ep
\]
We will condition on this event from now on and regard $Y_1,\dots,Y_n$ as constants. Note that $(X_1,\dots,X_n)$ and $(Y_1,\dots,Y_n)$ are independent.
   
Let $\delta\in (0,1/2)$ and introduce a $\delta$-net $S_\delta$ on the unit ball $\norm{u}{} \leq 1$. We have $\text{card}(S_\delta)\leq (1+\frac{2}{\delta})^{d}$ by~\citet[Corollary 4.2.13]{vershynin2018high}. By Lemma \ref{lem: Hoeffding}, for each $\norm{u}{}\leq 1$ and every $\ep>0$, we have \[
    \PP\left(\left|\frac{1}{n}\sum_{i=1}^{n}f(u^{\top}X_i)Y_i-\frac{1}{n}\sum_{i=1}^n \EE_{Z\sim \cN(0,1)}\left[f(\norm{u}{}Z)\right]Y_i\right| \geq \ep\right) \leq 2\operatorname{exp}\left(-c\frac{\ep^2n^2}{\sum_i Y_i^2}\right) .
    \]
Furthermore, doing union bound over $S_{\delta}$, we have
\[
\begin{aligned}
\PP\left(\operatorname{sup}_{u\in S_{\delta}}\left|\frac{1}{n}\sum_{i=1}^{n}f(u^{\top}X_i)Y_i-\frac{1}{n}\sum_{i=1}^n \EE_{Z\sim \cN(0,1)}\left[f(\norm{u}{}Z)\right]Y_i\right| \geq \ep\right) &\leq 2\left(1+\frac{2}{\delta}\right)^d\operatorname{exp}\left(-c\frac{\ep^2n^2}{\sum_i Y_i^2}\right)\\
& \leq 2\left(1+\frac{2}{\delta}\right)^d \operatorname{exp}(-c\ep^2n) .
\end{aligned}
\]
Now for any $\norm{u}{} \leq 1$, there exists $v\in S_\delta$ such that $\norm{v-u}{}\leq \delta$, and then 
    \begin{align*}
        \left| \frac{1}{n}\sum_{i=1}^{n}\left(f(u^{\top}X_i)-f(v^{\top}X_i)\right)Y_i \right|
        &\leq \frac{L}{n}\sum_{i=1}^{n}|u^{\top}X_i-v^{\top}X_i|\abs{Y_i}\\
        &\leq L\sqrt{\frac{1}{n}\sum_{i=1}^{n}|u^{\top}X_i-v^{\top}X_i|^2}\sqrt{\frac{1}{n}\sum_{i=1}^{n}Y_i^2}\\
        & \lesssim \delta L
    \end{align*}
    \wp at least $1-Ce^{Cd-cn}$,
    where in the last step we use Lemma \ref{lem: operator norm stronger} to uniformly control $\frac{1}{n}\sum_i \abs{u^{\top}X_i-v^{\top}X_i}^2$. Therefore, choosing $\delta=\ep/L$, we have 
    \[
    \begin{aligned}
&\abs{\frac{1}{n}\sum_{i=1}^n f(u^{\top}X_i)Y_i -\EE_{Z\sim \cN(0,1)}\left[f(\norm{u}{}Z)Y\right]}\\
& \quad\quad \leq \abs{\frac{1}{n}\sum_{i=1}^n f(u^{\top}X_i)Y_i-\frac{1}{n}\sum_{i=1}^n f(v^{\top}X_i)Y_i}+\abs{\frac{1}{n}\sum_{i=1}^n f(v^{\top}X_i)Y_i-\frac{1}{n}\sum_{i=1}^n \EE_{Z\sim \cN(0,1)}\left[f(\norm{v}{}Z)\right]Y_i}\\
&\quad\quad\quad+\abs{\frac{1}{n}\sum_{i=1}^n \EE_{Z\sim \cN(0,1)}\left[f(\norm{v}{}Z)\right]Y_i-\EE_{Z\sim \cN(0,1)}\left[f(\norm{v}{}Z)Y\right]} +C\delta L\\
& \quad\quad \leq C\delta L + \ep + \abs{\EE_{Z\sim \cN(0,1)}\left[f(\norm{v}{}Z)\right]}\ep \leq C\ep
    \end{aligned}
    \]
    \wp at least $1-Ce^{Cd-cn}-2(1+2L/\ep)^d\operatorname{exp}(-c\ep^2 n)$.
\end{proof}

\begin{lemma}\label{lem: III concentrate part 3}
    Suppose $f:\RR\to\RR$ is a locally Lipschitz continuous function such that 
    \[|f(x)-f(\Tilde{x})|\leq L (1+|x|+|\Tilde{x}|)|x-\Tilde{x}|,\quad \forall x,\Tilde{x}\in\RR.\]
    Assume that $\sup_{0\leq a \leq 1}\|f(aZ)\|_{\psi_1}\leq C$ for a standard Gaussian $Z\sim \cN(0,1)$.
    
        Assume $X_1,\dots,X_n$ are \iid $\cN(0,I_d)$ generated random variables.
    Assume that $Y_1,\dots,Y_n$ are \iid random variables with $\EE\left[Y^4\right] \leq C$ that is independent of the $X_1,\dots,X_n$. Further assume $0\leq Y\leq M$ almost surely.
    
    For any $\ep\in (0,1)$, it holds 
    \[\sup_{\norm{u}{} \leq 1}\left|\frac{1}{n}\sum_{i=1}^{n}f(u^{\top}X_i)Y_i-\EE_{Z\sim \cN(0,1)}\left[ f(\norm{u}{}Z)Y\right]\right|\leq C\ep
    \]
    \wp at least $(1-Ce^{Cd-cn/M}-2\left(1+2L/\ep\right)^d\operatorname{exp}\left(-c\operatorname{min}(n\ep^2,n\ep/M)\right))(1-C\frac{1}{\ep^2n})$.
\end{lemma}
\begin{proof}

Using standard Markov's inequality arguments, we have \wp at least $1-C\frac{1}{\ep^2 n}$,
\[
\abs{\frac{1}{n}\sum_{i=1}^n Y_i-\EE\left[Y\right]} \leq \ep \text{  and  } \abs{\frac{1}{n}\sum_{i=1}^n Y_i^2 -\EE\left[Y^2\right]} \leq \ep .
\]
We will condition on this event from now on and regard $Y_1,\dots, Y_n$ as constants. Note that $(X_1,\dots, X_n)$ and $(Y_1,\dots,Y_n)$ are independent.
    
    Let $\delta\in (0,1/2)$ and introduce a $\delta$-net $S_\delta$ on $\norm{u}{}\leq 1$. We have $\text{card}(S_\delta)\leq (1+\frac{2}{\delta})^{d}$ by~\citet[Corollary 4.2.13]{vershynin2018high}. 
First, by Bernstein inequality, Lemma \ref{lem: Berstein}, for each $\norm{u}{}\leq 1$, we have 
\[
    \PP\left(\left|\frac{1}{n}\sum_{i=1}^{n}f(u^{\top}X_i)Y_i-\frac{1}{n}\sum_{i=1}^n \EE_{Z\sim \cN(0,1)}\left[f(\norm{u}{}Z)\right]Y_i\right| \geq \ep\right) \leq 2\operatorname{exp}\left(-c\operatorname{min}(n\ep^2,n\ep/M)\right) .
    \]
Furthermore, doing union bound over $S_{\delta}$, we have
\[
\begin{aligned}
\PP\left(\operatorname{sup}_{u\in S_{\delta}}\left|\frac{1}{n}\sum_{i=1}^{n}f(u^{\top}X_i)Y_i-\frac{1}{n}\sum_{i=1}^n \EE_{Z\sim \cN(0,1)}\left[f(\norm{u}{}Z)\right]Y_i\right| \geq \ep\right) &\leq 2\left(1+\frac{2}{\delta}\right)^d\operatorname{exp}\left(-c\operatorname{min}(n\ep^2,n\ep/M)\right) .
\end{aligned}
\]
  Now for any $\norm{u}{}\leq 1$, there exists $v\in S_\delta$ such that $\norm{u-v}{}\leq \delta$, and then 
    \begin{align*}
        &\left| \frac{1}{n}\sum_{i=1}^{n}\left(f(u^{\top}X_i)-f(v^{\top}X_i)\right)Y_i \right|
        \lesssim \frac{L}{n}\sum_{i=1}^{n}|u^{\top}X_i-v^{\top}X_i|(1+|u^{\top}X_i|+|v^{\top}X_i|)Y_i\\
        &\quad\quad\quad\quad\lesssim L\sqrt{\frac{1}{n}\sum_{i=1}^{n}|u^{\top}X_i-v^{\top}X_i|^2Y_i}\left(\sqrt{\frac{1}{n}\sum_{i=1}^{n}|u^{\top}X_i|^2Y_i}+\sqrt{\frac{1}{n}\sum_{i=1}^{n}|v^{\top}X_i|^2Y_i}+C\right) .\\
    \end{align*}
By our Lemma \ref{lem: operator norm stronger}, \wp at least $1-Ce^{Cd-cn/M}$, we have
\[
\frac{1}{n}\sum_{i=1}^n (u^{\top}X_i)^2 Y_i \leq C
\]
uniformly for all $\norm{u}{}\leq 1$. Therefore, \wp at least $1-Ce^{Cd-cn/M}$, we have\[
\left| \frac{1}{n}\sum_{i=1}^{n}\left(f(u^{\top}X_i)-f(v^{\top}X_i)\right)Y_i \right| \lesssim \delta L .
\]
Therefore, choosing $\delta=\ep/L$, we have
\[
\begin{aligned}
&\abs{\frac{1}{n}\sum_{i=1}^n f(u^{\top}X_i)Y_i -\EE_{Z\sim \cN(0,1)}\left[f(\norm{u}{}Z)Y\right]}\\
& \quad\quad \leq \abs{\frac{1}{n}\sum_{i=1}^n f(u^{\top}X_i)Y_i-\frac{1}{n}\sum_{i=1}^n f(v^{\top}X_i)Y_i}+\abs{\frac{1}{n}\sum_{i=1}^n f(v^{\top}X_i)Y_i-\frac{1}{n}\sum_{i=1}^n \EE_{Z\sim \cN(0,1)}\left[f(\norm{v}{}Z)\right]Y_i}\\
&\quad\quad\quad+\abs{\frac{1}{n}\sum_{i=1}^n \EE_{Z\sim \cN(0,1)}\left[f(\norm{v}{}Z)\right]Y_i-\EE_{Z\sim \cN(0,1)}\left[f(\norm{v}{}Z)Y\right]} +C\delta L\\
& \quad\quad \leq C\delta L + \ep + \abs{\EE_{Z\sim \cN(0,1)}\left[f(\norm{v}{}Z)\right]}\ep \lesssim \ep
    \end{aligned}
\]
\wp at least $1-Ce^{Cd-cn/M}-2\left(1+2/\ep\right)^d\operatorname{exp}\left(-c\operatorname{min}(n\ep^2,n\ep/M)\right)$.
\end{proof}
\begin{lemma}\label{lem: lip concentrate}
  Suppose $f:\RR\to\RR$ is a Lipschitz continuous function such that 
    \[|f(x)-f(\Tilde{x})|\leq L |x-\Tilde{x}|,\quad \forall x,\Tilde{x}\in\RR.\]
    Suppose $f$ has compact support $[-N,N]$.
Denote $X\sim \cN(0,I_d)$ and $X_1,\dots,X_n$ are \iid generated samples. Denote $Y_1,\dots,Y_n$ are \iid generated samples and assume $\norm{Y}{\infty} \leq t$ and $\EE\left[Y^2\right]\leq C$.
    
   Then, for any $\ep>0$, it holds
    \[\sup_{u\in \SS^{d-1}}\left|\frac{1}{n}\sum_{i=1}^{n}f(u^{\top}X_i)Y_i-\EE \left[f(u^{\top}X)Y\right]\right| \leq C\ep\]
    w.p. at least $1-Ce^{Cd-cn}-C/n-2(1+2L/\ep)^d \operatorname{exp}\left(-c\frac{n\ep^2}{L^2N^2t^2}\right)$.
\end{lemma}
\begin{proof}
    From the assumptions, we have $\norm{f}{\infty}\leq LN$.
    Therefore $f (u^{\top}X)Y$ is bounded by $LNt$, and we can apply Lemma \ref{lem: Hoeffding's inequality for bounded} and obtain the following inequality for each $u\in\SS^{d-1}$:
    \[\PP\left(\left|\frac{1}{n}\sum_{i=1}^{n}f(u^{\top}X_i)Y_i-\EE \left[f(u^{\top}X)Y\right]\right|\geq \ep\right) \leq 2\operatorname{exp}\left(-c\frac{n\ep^2}{L^2N^2t^2}\right) .
    \]
    Let $\delta\in (0,1/2)$ and introduce a $\delta$-net $S_\delta$ on $\SS^{d-1}$. We have $\text{card}(S_\delta)\leq (1+\frac{2}{\delta})^d$ by~\citet[Corollary 4.2.13]{vershynin2018high}. Therefore, taking a union bound, we have
    \[
    \PP\left(\sup_{u\in S_{\delta}}\left|\frac{1}{n}\sum_{i=1}^{n}f(u^{\top}X_i)Y_i-\EE \left[f(u^{\top}X)Y\right]\right|\geq \ep\right) \leq 2\left(1+\frac{2}{\delta}\right)^d\operatorname{exp}\left(-c\frac{n\ep^2}{L^2N^2t^2}\right) .
    \]
    For any $u\in \SS^{d-1}$, there exists $v\in S_\delta$ such that $\norm{u-v}{}\leq \delta$, and then 
    \begin{align*}
        \left|\frac{1}{n}\sum_{i=1}^{n}\left(f(u^{\top}X_i)-f(v^{\top}X_i)\right)Y_i\right|
        &\leq \frac{1}{n}\sum_{i=1}^{n}\left|f(u^{\top}X_i)-f(v^{\top}X_i)\right|\abs{Y_i}\\
        &\leq \frac{L}{n}\sum_{i=1}^{n}\left|u^{\top}X_i-v^{\top}X_i\right|\abs{Y_i}\\
        &\leq L \sqrt{\frac{1}{n}\sum_{i=1}^{n}\left|u^{\top}X_i-v^{\top}X_i\right|^2}\sqrt{\frac{1}{n}\sum_{i=1}^{n}Y_i^2}\\
        &\lesssim L\delta
    \end{align*}
\wp at least $1-Ce^{Cd-cn}-C/n$, due to standard Markov's inequality arguments and Lemma \ref{lem: operator norm stronger}.
    It is also not difficult to control the differences in expectation.
    \begin{align*}
        \left|\EE\left[ \left(f(u^{\top}X)-f(v^{\top}X)\right)Y\right]\right|
        &\leq \EE \left[\left|f(u^{\top}X)-f(v^{\top}X)\right|\abs{Y}\right]\\
        &\leq L\EE\left[\left|u^{\top}X-v^{\top}X\right|\abs{Y}\right]\\
        &\leq L \sqrt{\EE\left[\left|u^{\top}X-v^{\top}X\right|^2\right]}\sqrt{\EE\left[ Y^2\right]}\\
        &\lesssim L \delta .
    \end{align*}
Therefore, set $\delta = \ep/L$, we have the following for uniformly all $u\in \SS^{d-1}$
\[
\begin{aligned}
    \left|\frac{1}{n}\sum_{i=1}^{n}f(u^{\top}X_i)Y_i-\EE \left[f(u^{\top}X)Y\right]\right|
    &\leq \left|\frac{1}{n}\sum_{i=1}^{n}\left(f(u^{\top}X_i)-f(v^{\top}X_i)\right)Y_i\right|\\
    &\quad+ \left|\frac{1}{n}\sum_{i=1}^{n}f(v^{\top}X_i)Y_i-\EE \left[f(v^{\top}X)Y\right]\right|
    + \left|\EE \left[\left(f(u^{\top}X)-f(v^{\top}X)\right)Y\right]\right|\\
    & \leq C\ep
\end{aligned}
\]
\wp at least $1-Ce^{Cd-cn}-C/n-2(1+2L/\ep)^d \operatorname{exp}\left(-c\frac{n\ep^2}{L^2N^2t^2}\right)$.
\end{proof}

\section{Technical Toolbox}\label{appendix: technical}

\subsection{Classical Concentration Inequalities}
\begin{definition}[Sub-exponential Random Variable]
    For a random variable $X$, define \begin{equation}
\|X\|_{\psi_1}=\inf \{t>0: \EE \exp (|X| / t) \leq 2\}
\end{equation}
to be its sub-exponential norm. A random variable $X$ is said to be sub-exponential if  $\norm{X}{\psi_1}< +\infty$.
\end{definition}
\begin{definition}[Sub-gaussian Random Variable]
    For a random variable $X$, define \begin{equation}
\|X\|_{\psi_2}=\inf \{t>0: \EE \exp (X^2 / t^2) \leq 2\}
\end{equation}
to be its sub-gaussian norm. A random variable $X$ is said to be sub-gaussian if  $\norm{X}{\psi_2}< +\infty$.
\end{definition}

We will frequently use the following   standard concentration inequalities for bounded, sub-exponential and sub-gaussian random variables.
\begin{theorem}(\citep[Theorem 2.2.6]{vershynin2018high})\label{lem: Hoeffding's inequality for bounded}
    Let $X_1,\dots,X_n$ be independent random variables. Assume that $X_i\in [m_i,M_i]$ for every $i\in [n]$. Then, for any $t>0$, we have 
    \[\PP\left(\abs{\sum_{i=1}^n X_i-\EE \left[X_i\right]}\geq t\right)\leq 2\exp{\left[-\frac{2t^2}{\sum_{i=1}^n (M_i-m_i)^2}\right]}.\]
\end{theorem}
\begin{theorem}(\citep[Theorem 2.8.2]{vershynin2018high})\label{lem: Berstein}
    Let $X_1.\dots,X_n$ be independent, mean zero, sub-exponential random variables, and $a=(a_1,\dots,a_n)\in\RR^n$. Then, for any $t>0$, we have
    \[\PP\left(\left|\sum_{i=1}^na_iX_i\right|\geq t\right)\leq 2\exp{\left[-c\min\left(\frac{t^2}{K^2\|a\|^2},\frac{t}{K\|a\|_\infty}\right)\right]}\]
    where $K=\max_i\norm{X_i}{\psi_1}$.
\end{theorem}

\begin{theorem}(\citep[Theorem 2.6.3]{vershynin2018high})\label{lem: Hoeffding}
    Let $X_1,\dots,X_n$ be independent, mean zero, sub-gaussian random variables and $a=(a_1,\cdots,a_n)\in\RR^n$. Then for every $t>0$, we have
    \[\PP\left(\left|\sum_{i=1}^n a_iX_i\right|\geq t\right)\leq 2\exp{\left(-c\frac{t^2}{K^2\|a\|^2}\right)}\]
    where $K=\max_i\norm{X_i}{\psi_2}$.
\end{theorem}

\subsection{Results for Gaussian Random Variables}
\begin{lemma}\label{lemma: eigen decomp for matrix}
    Consider a $2\times 2$ symmetric matrix $V=\begin{pmatrix}
 a & b\\
 b & d
\end{pmatrix}$.
Then there exists an orthogonal matrix $U$ such that $U^{\top}VU$ is diagonal.
\end{lemma}
\begin{proof}
By solving 
$
    \mathrm{det}(\lambda I - V) = (\lambda-a)(\lambda-d) -b^2=0,
$
we have the eigenvalues of $V$ given by 
\begin{equation}\label{eqn: lambda_2x2}
\lambda_\pm = \frac{a+d\pm \sqrt{(a-d)^2+4b^2}}{2}
\end{equation}
and  eigenvectors $U$ can explicitly constructed as follows
    \[U=\begin{pmatrix}
\frac{b}{\sqrt{b^2+(\lambda _+-a)^2} } & \frac{b}{\sqrt{b^2+(\lambda _--a)^2} }\\
\frac{\lambda_+-a}{\sqrt{b^2+(\lambda _+-a)^2}}  & \frac{\lambda_--a}{\sqrt{b^2+(\lambda _--a)^2}}.
\end{pmatrix} \] 
Then, it is easy to  verify that $U^{\top}VU=\Lambda$ where $\Lambda=\begin{pmatrix}
 \lambda _+ & \\
  & \lambda _-
\end{pmatrix}$.
\end{proof}

\begin{lemma}
(Tails of the normal distribution,~\citep[Proposition 2.1.2]{vershynin2018high})\label{lemma: tail of gaussian} Let $X \sim \cN(0,1)$. Then for all $t> 0$, we have
$$
\left(\frac{1}{t}-\frac{1}{t^3}\right) \cdot \frac{1}{\sqrt{2 \pi}} e^{-t^2 / 2} \leq \PP\{X \geq t\} \leq \frac{1}{t} \cdot \frac{1}{\sqrt{2 \pi}} e^{-t^2 / 2} .
$$
\end{lemma}

\begin{lemma}\label{lemma: lower exp tail for quadratic form}

Consider a quadratic form $f(x,y)=ax^2+2bxy+dy^2$ with $a>0$, $d>0$ and $b^2-ad>0$. Further assume $a,\abs{b},d\leq C$. Let $X_1,X_2$ be two independent $\cN(0,1)$ random variables. Then, for any $t\geq C$, we have
        \[
        \PP\left\{f(X_1,X_2) \leq -t\right\} \gtrsim \sqrt{\frac{-\lambda_{-}}{t}}e^{t/\lambda_{-}}e^{\lambda_+/\lambda_{-}},
        \]
where $\lambda_{\pm}$ is given by Eq.~\eqref{eqn: lambda_2x2}.
\end{lemma}
The condition $b^2-ad>0$ ensures $\lambda_{-}<0$.
\begin{proof}
Denote $X=(X_1,X_2)^{\top}$. Applying Lemma \ref{lemma: eigen decomp for matrix} along with the same notation, we have
\[
f(X_1,X_2)=X^{\top}VX=X^{\top}U\Lambda U^{\top}X
\]
Let $Y=U^{\top}X=(Y_1,Y_2)^{\top}$. Then $Y_1,Y_2$ are independent $\cN(0,1)$ random variables and $f(X_1,X_2)=Y^{\top}\Lambda Y= \lambda_{+}Y_1^2 +\lambda_{-}Y_2^2$. When $b^2-ad>0$, we have $\lambda_{-}<0$, and we can do the following estimate for the tail of $f(X_1,X_2)$:
\[
\begin{aligned}
\PP\left(f(X_1,X_2) \leq -t\right) &= \PP\left(\lambda_{+}Y_1^2 +\lambda_{-}Y_2^2 \leq -t\right)=\PP\left(Y_2^2 \geq \frac{t+\lambda_{+}Y_1^2}{-\lambda_{-}}\right)\\
& \gtrsim \PP\left(Y_2^2 \geq \frac{t+\lambda_{+}Y_1^2}{-\lambda_{-}}, Y_1^2\leq 1\right)\gtrsim \PP\left(Y_2^2 \geq \frac{t+\lambda_{+}}{-\lambda_{-}}, Y_1^2\leq 1\right) \\ 
&=\PP\left(Y_2^2 \geq \frac{t+\lambda_{+}}{-\lambda_{-}}\right)\PP\left(Y_1^2\leq 1\right)\gtrsim \PP\left(Y_2^2 \geq  \frac{t+\lambda_+}{-\lambda_{-}}\right) \\
& \stackrel{\text{Lemma \ref{lemma: tail of gaussian}}}{\gtrsim} \left(\sqrt{\frac{-\lambda_-}{t+\lambda_+}}-\left(\frac{-\lambda_-}{t+\lambda_+}\right)^{3/2}\right)e^{\frac{t+\lambda_+}{2\lambda_-}}\\
& \gtrsim \sqrt{\frac{-\lambda_{-}}{t}}e^{\frac{t}{2\lambda_{-}}}e^{\frac{\lambda_+}{2\lambda_{-}}},
\end{aligned}
\]
where the last inequality uses that $t$ is sufficiently  large compared to $\lambda_{+}$.
\end{proof}

\begin{lemma}(\citep[Theorem 3.1.1]{vershynin2018high})\label{lemma: concetration of the norm}
    Suppose $Z\sim \cN(0,I_d)$, then w.p. at least $1-Ce^{-ct^2}$ we have
    \begin{equation*}
        \abs{\norm{Z}{}-\sqrt{d}}\leq t .
    \end{equation*}
    
\end{lemma}

\begin{lemma}(Borell-TIS Inequality,~\citep[Theorem 2.1.1]{adler2007random})\label{lemma: gaussian extreme concentration}
    Let $X_1,\dots,X_n$ be \iid $\cN(0,1)$ random variables and denote $X_{\operatorname{min}}:=\operatorname{min}_i X_i$. Then for each $t>0$,
    \[\PP(|X_{\operatorname{min}}-\EE\left[X_{\operatorname{min}}\right]|\geq t)\leq 2 e^{-t^2/2} .\]
\end{lemma}

\begin{lemma}(\citep[Exercise 2.5.10 and 2.5.11]{vershynin2018high})\label{lemma: expectation of gaussian extreme}
    Let $X_1,\dots,X_n$ be \iid $\cN(0,1)$  random variables and denote $X_{\operatorname{min}}:=\operatorname{min}_i X_i$. Then for every $n\in \NN$,
\[
\abs{\EE\left[X_{\operatorname{min}}\right]} = \Theta\left(\sqrt{\log n}\right) .
\]
\end{lemma}
\begin{proof}

    Exercise 2.5.10 and 2.5.11 in \cite{vershynin2018high} together show that $\EE[\max_{i\in [n]}X_i]= \Theta(\sqrt{\log n})$ for $X_i\stackrel{iid}{\sim}\cN(0,1)$. Noticing $X_i$ and $-X_i$ follow the same distribution. Therefore, 
    \[
        \EE[\min_{i\in [n]}X_i] = - \EE[\max_{i\in [n]}X_i] = -\Theta(\sqrt{\log n}).
    \]
    
\end{proof}

\begin{lemma}\label{lem: moment cutoff}
    Let $X\sim\cN(0,1)$. Then for all $t>0$, we have 
    \[t\cdot \frac{2}{\sqrt{2\pi}}e^{-\frac{t^2}{2}}\leq \EE\left[ X^2 1_{|X|\geq t}\right]\leq (t+\frac{1}{t})\cdot \frac{2}{\sqrt{2\pi}}e^{-\frac{t^2}{2}},\]
    \[(t^3+3t)\cdot \frac{2}{\sqrt{2\pi}}e^{-\frac{t^2}{2}}\leq \EE\left[ X^41_{|X|\geq t}\right]\leq (t^3+3t+\frac{3}{t})\cdot \frac{2}{\sqrt{2\pi}}e^{-\frac{t^2}{2}},\]
    \[(t^7+7t^5+35t^3+105t)\cdot \frac{2}{\sqrt{2\pi}}e^{-\frac{t^2}{2}}\leq \EE \left[X^81_{|X|\geq t}\right]\leq (t^7+7t^5+35t^3+105t+\frac{105}{t})\cdot \frac{2}{\sqrt{2\pi}}e^{-\frac{t^2}{2}}.\]
\end{lemma}
\begin{proof}
    For any even integer $n\geq 2$, we have 
    \begin{align*}
    \EE \left[X^n1_{|X|\geq t}\right] &= \frac{2}{\sqrt{2\pi}}\int_t^\infty x^n e^{-x^2/2}\dd x=\frac{2}{\sqrt{2\pi}}\int_t^\infty x^{n-1} (-e^{-x^2/2})'\dd x\\ 
    &= \frac{2}{\sqrt{2\pi}} \left(-e^{-x^2/2}\cdot x^{n-1}\Big|_{t}^\infty - \int_t^\infty e^{-x^2/2}\cdot (n-1) x^{n-2}\dd x\right)\\ 
    &=  \frac{2}{\sqrt{2\pi}}e^{-\frac{t^2}{2}}t^{n-1} + (n-1)\EE[X^{n-2}1_{|X|\geq t}].
    \end{align*}
     Specifically, we have
\[\EE \left[X^21_{|X|\geq t}\right]=t\cdot \frac{2}{\sqrt{2\pi}}e^{-\frac{t^2}{2}} + 2\PP(X\geq t),\]
    \[\EE \left[X^41_{|X|\geq t}\right]=(t^3+3t)\cdot \frac{2}{\sqrt{2\pi}}e^{-\frac{t^2}{2}} + 6\PP(X\geq t),\]
    \[\EE \left[X^61_{|X|\geq t}\right]=(t^5+5t^3+15t)\cdot \frac{2}{\sqrt{2\pi}}e^{-\frac{t^2}{2}} + 30\PP(X\geq t),\]
    \[\EE \left[X^81_{|X|\geq t}\right]=(t^7+7t^5+35t^3+105t)\cdot \frac{2}{\sqrt{2\pi}}e^{-\frac{t^2}{2}} + 210\PP(X\geq t).\]
    Lastly, plugging Lemma \ref{lemma: tail of gaussian} in, we complete the proof.
\end{proof}
\begin{lemma}\label{lem: density concentration}
    Let $X$ and $X_1,\dots,X_n$ be \iid $\cN(0,1)$ random variables. Assume $t \geq 1$. Then for any $\ep>0$, w.p. at least $1-C\frac{t}{\ep^2 n}e^{\frac{t^2}{2}}$, we have\[\left|\frac{1}{n}\sum_{i=1}^{n}X_i^41_{|X_i|\geq t} - \EE \left[X^41_{|X|\geq t}\right]  \right|\leq \ep\EE \left[X^41_{|X|\geq t}\right].\]
\end{lemma}
\begin{proof}
By Lemma \ref{lem: moment cutoff}, we have when $t$ is sufficiently large,
   $\EE \left[X^41_{|X|\geq t}\right] = \Theta(t^3e^{-\frac{t^2}{2}})$ and $\EE \left[X^81_{|X|\geq t}\right] = \Theta(t^7e^{-\frac{t^2}{2}})$.
By Chebyshev's inequality, we have
\begin{align*}
    \PP\left(\left|\frac{1}{n}\sum_{i=1}^{n}X_i^41_{|X_i|\geq t} - \EE \left[X^41_{|X|\geq t}\right]  \right|\geq \ep\EE \left[X^41_{|X|\geq t}\right]\right) & \leq \frac{\operatorname{Var}\left(\frac{1}{n}\sum_{i=1}^{n}X_i^41_{|X_i|\geq t}\right)}{\ep^2\left(\EE\left[X^4 1_{\abs{X}\geq t}\right]\right)^2}\\
    &\leq \frac{\EE \left[X^81_{|X|\geq t}\right]}{\ep^2n\left(\EE\left[X^4 1_{\abs{X}\geq t}\right]\right)^2}\\
    &\lesssim \frac{t}{\ep^2 n}e^{\frac{t^2}{2}}.
\end{align*}
Then, the conclusion follows.
\end{proof}

\begin{lemma}\label{lem: gaussian inner product}
    Suppose $Z_i\stackrel{iid}{\sim} \cN(0,I_d)$ for $i=1,\dots,n+1$ and let $U_i := Z_i^{\top}\frac{Z_{n+1}}{\norm{Z_{n+1}}{}}$ for $i=1,\cdots,n$. Then $U:=(U_1,\cdots,U_n)^\top\sim \cN(0,I_n)$. 
\end{lemma}
\begin{proof}
    We prove this lemma by verifying the characteristic function of $U$.
     Noting  for any fix $Z_{n+1}$, $Z_i^\top \frac{Z_{n+1}}{\norm{Z_{n+1}}{}}\stackrel{iid}{\sim}\cN(0,1)$, we have for $t=(t_1,\cdots,t_n)^\top\in\RR^n$, it holds that
    \begin{align*}
        \EE e^{it^\top U}&=\EE e^{i\sum_{i=1}^n t_iU_i}\\
        &=\EE \left[\EE \left[e^{i\sum_{i=1}^n t_iU_i}| Z_{n+1}\right]\right]\\
        &=\EE \left[\EE \left[e^{i\sum_{i=1}^n t_iZ_i^\top \frac{Z_{n+1}}{\norm{Z_{n+1}}{}}}\Big| Z_{n+1}\right]\right]\\
        &=\EE [e^{-\frac{1}{2}\sum_{i=1}^nt_i^2}| Z_{n+1}]\\
        &=e^{-\frac{1}{2}\sum_{i=1}^nt_i^2},
    \end{align*}
    which is identical to the characteristic function of  $\cN(0,I_n)$. Thus, we complete the proof.
\end{proof}


\end{document}